\documentclass[12pt,subeqn]{article}

\usepackage{amsmath}

\renewcommand{\theequation}{\arabic{equation}}
\newcommand{\tr}{\operatorname{Tr}}

\newcommand{\EQ}{\begin{equation}}
\newcommand{\EN}{\end{equation}}

\newcommand{\ket}[1]{\left|#1\right\rangle}      % Ket-Zustand
\newcommand{\bear}{\begin{eqnarray}}
\newcommand{\ear}{\end{eqnarray}}
\newcommand{\bt} { \begin{tabular} }
\newcommand{\et}{ \end{tabular} }
\newcommand{\bc} { \begin{center} }
\newcommand{\ec}{ \end{center} }

\newcommand{\btb} { \begin{table} }
\newcommand{\etb}{ \end{table} }
\begin{document}

\topmargin 0pt
\oddsidemargin 5mm
\newcommand{\NP}[1]{Nucl.\ Phys.\ {\bf #1}}
\newcommand{\PL}[1]{Phys.\ Lett.\ {\bf #1}}
\newcommand{\NC}[1]{Nuovo Cimento {\bf #1}}
\newcommand{\CMP}[1]{Comm.\ Math.\ Phys.\ {\bf #1}}
\newcommand{\PR}[1]{Phys.\ Rev.\ {\bf #1}}
\newcommand{\PRL}[1]{Phys.\ Rev.\ Lett.\ {\bf #1}}
\newcommand{\MPL}[1]{Mod.\ Phys.\ Lett.\ {\bf #1}}
\newcommand{\JETP}[1]{Sov.\ Phys.\ JETP {\bf #1}}
\newcommand{\TMP}[1]{Teor.\ Mat.\ Fiz.\ {\bf #1}}

\renewcommand{\thefootnote}{\fnsymbol{footnote}}

\newpage
\setcounter{page}{0}
\begin{titlepage}
\begin{flushright}
UFSCARF-TH-2009-10
\end{flushright}
\vspace{0.5cm}
\begin{center}
{\large Algebraic Bethe Ansatz for $U(1)$ Invariant Integrable Models: \\
       Compact and non-Compact Applications }\\
\vspace{1cm}
{\large M.J. Martins and C.S. Melo} \\
\vspace{0.5cm}
{\em Universidade Federal de S\~ao Carlos\\
Departamento de F\'{\i}sica \\
C.P. 676, 13565-905~~S\~ao Carlos(SP), Brasil}\\
\end{center}
\vspace{0.5cm}

\begin{abstract}
We apply the algebraic Bethe ansatz developed in our previous paper \cite{CM}
to three different families of 
$U(1)$ integrable vertex models with arbitrary $N$ bond states. 
These statistical mechanics systems are based on the
higher spin representations of the 
quantum group $U_q[SU(2)]$ for both generic 
and non-generic values of
$q$ as well as
on the non-compact discrete representation 
of the $SL(2,{\cal R})$ algebra. 
We present for all these models
the explicit expressions for both
the on-shell and the off-shell properties associated to the 
respective transfer matrices eigenvalue problems. 
The amplitudes governing the vectors not parallel to the Bethe states are shown to
factorize 
in terms of elementary building blocks
functions. The results for the 
non-compact $SL(2,{\cal R})$ model are argued to be derived from those obtained for the compact
systems by taking suitable
$N \rightarrow \infty$ limits. This permits us 
to study the properties of the non-compact
$SL(2,{\cal R})$ model starting from systems with finite degrees of freedom.
\end{abstract}
\vspace{.09cm}
\centerline{PACS numbers:  05.50+q, 02.30.IK}
\vspace{.02cm}
\centerline{Keywords: Algebraic Bethe Ansatz, Lattice Integrable Models}
\vspace{.01cm}
\centerline{March 2009}

\end{titlepage}

%\tableofcontents

\pagestyle{empty}

\newpage

\pagestyle{plain}
\pagenumbering{arabic}

\renewcommand{\thefootnote}{\arabic{footnote}}

\section{Introduction}

This article is a continuation of the paper \cite{CM} in which we have developed 
the algebraic Bethe ansatz for $U(1)$ invariant vertex models. 
The central object
in this method turns out to be the monodromy matrix 
${\cal T}_{\cal A}(\lambda) $ \cite{FA,KO} depending on the spectral
parameter $\lambda$. For a $N$-dimensional auxiliary space ${\cal A}$  we view 
${\cal T}_{\cal A}(\lambda) $ as,
\EQ
{\cal T_{A}}(\lambda)=\left(\begin{array}{cccc}
                {\cal T}_{1,1}(\lambda) & {\cal T}_{1,2}(\lambda) & \cdots & {\cal T}_{1,N}(\lambda) \\
                {\cal T}_{2,1}(\lambda) & {\cal T}_{2,2}(\lambda) & \cdots & {\cal T}_{2,N}(\lambda) \\
                \vdots & \vdots & \ddots & \vdots \\
                {\cal T}_{N,1}(\lambda) & {\cal T}_{N,2}(\lambda) & \cdots & {\cal{T}}_{N,N}(\lambda) \\
                \end{array}\right)_{N \times N},
\label{mono}
\EN
where the elements ${\cal T}_{a,b}(\lambda) $  are operators acting on the system quantum space.

The physical quantities such as partition function can be expressed in terms of the trace of
the monodromy matrix. This operator is the generating function of the conserved currents and is
called the transfer matrix $T(\lambda)$,
\EQ
T(\lambda)=\tr_{\cal A}\left[ {\cal T}_{\cal A}(\lambda) \right]=\sum_{i=1}^{N} {\cal{T}}_{a,a}(\lambda).
\label{tran}
\EN

A necessary condition to diagonalize $T(\lambda)$ by the algebraic Bethe ansatz is the existence
of a vector $\ket{0}$ such that the action of the monodromy on it results in a triangular matrix
for arbitrary values of the spectral parameter. For instance, if 
${\cal T}_{\cal A}(\lambda) \ket{0}$ is annihilated by its lower left elements we have,
\EQ
{\cal T_{A}}(\lambda) \ket{0}=\left(\begin{array}{cccc}
                \omega_1(\lambda) \ket{0} & \# & \cdots & \# \\
                0 & \omega_2(\lambda)\ket{0} & \cdots & \# \\
                \vdots & \vdots & \ddots & \vdots \\
                0 & 0 & \cdots & \omega_{N}(\lambda)\ket{0} \\
                \end{array}\right)_{N \times N},
\label{action}
\EN
where the symbol $\#$ denotes non-null states and $\omega_i(\lambda)$ for $i=1,\cdots,N$ are
complex valued functions.

After having property (\ref{action}) fulfilled one can in principle use   
the algebraic Bethe ansatz method to 
propose an ansatz for 
other eigenvectors $\ket{\Phi_n}$ of $T(\lambda)$. In general, the
states $\ket{\Phi_n}$ are searched as linear 
combination of certain product of creation fields 
${\cal T}_{a,b}(\lambda) $ with $a <b$ acting on the reference state $\ket{0}$.
The next step in this framework is to compute the action of the diagonal operators 
${\cal T}_{a,a}(\lambda) $ on the ansatz state
$\ket{\Phi_n}$.  This operation generates not only the state
$\ket{\Phi_n}$ but also a number of vectors that are not parallel
to such proposed eigenvectors of $T(\lambda)$ which are often called unwanted terms.
This analysis is then able to produce informations on the
on-shell Bethe ansatz properties which turns out to be the 
eigenvalues of $T(\lambda)$ and the Bethe ansatz equations needed
to cancel out the unwanted terms. In addition, it also gives us the 
off-shell properties which consist on both the
determination of the 
pattern of the vectors not parallel
to $\ket{\Phi_n}$ and the functional form of the functions proportional to
these states. We shall refer to these functions as the off-shell
amplitudes of a given algebraic Bethe ansatz analysis. We remark that the 
knowledge of the off-shell Bethe ansatz data is of relevance since its
semi-classical limit can
provide solutions of integrable long-range systems such as
the Gaudin models \cite{GA}  as well as  representations for 
the solutions of equations of
Knizhnik-Zamolodchikov type \cite{BAB}.

In our previous paper \cite{CM} we have developed the above discussed algebraic
framework to solve arbitrary integrable vertex models that are invariant
by one $U(1)$ symmetry. Here we will use the general results of \cite{CM}
to compute the explicit expressions for both the on-shell and off-shell
parts of the algebraic Bethe ansatz solution of three distinct classes of
vertex models. We recall that the off-shell properties have been given
in terms of recurrence relations whose solution for a specific model requires
additional analysis.  
At first we consider the vertex model derived from the higher spin
representation of the $U_q[SU(2)]$ algebra for generic values of
the deformation parameter $q$. This gives origin to the celebrated 
integrable spin-${\bf s}$ extension of the 
Heisenberg XXZ chain \cite{FA1,SO,RKS,FAT}.
The on-shell properties of this system have been obtained long ago
by using the fusion hierarchy procedure \cite{BAB1,RK}. 
For a recent discussion
on the algebraic Bethe ansatz of the XXZ-${\bf s}$ in the context of
the super-integrable chiral Potts model see \cite{DEG2}. However, to the best of our
knowlodge the off-shell Bethe ansatz
structure of the XXZ-${\bf s}$ is still unknown. The second family of models we shall consider
are those directly associated to the colored solutions of 
the Yang-Baxter equation \cite{DEG,DEG3,DEG1}. These vertex models are intimately
connected to the representations of the $U_q[SU(2)]$ algebra when
$q$ is a root of unity \cite{COT1,SIE}. The third system is based on the discrete
$D_{{\bf s}}^{-}$ representation of the 
$SL(2,{\cal R})$ symmetry leading us to a vertex model with an infinite number
of states per bond.

We have organized this paper as follows. For sake of completeness
we review the main results of our previous work \cite{CM} in next section.
This helps us to elaborate on  the on-shell
and off-shell Bethe ansatz results discussed in \cite{CM} as well as to
present them in a self-consistent way. In section 
3 we consider the
classical analogue of the solvable spins-${\bf s}$ XXZ model. Its
$R$-matrix is expressed in the Weyl basis in order to allow us
to compute the respective off-shell properties in closed
forms. In section 4 we discuss the vertex models derived from the
braid group representations associated to the $U_q[SU(2)]$ quantum
algebra when $q$ is a root of unity. The main feature of this system
is that its Boltzmann weights may depend on three
distinct continuous variables and this freedom is used to
define vertex models with additive and non-additive $R$-matrices. 
The corresponding on-shell and off-shell Bethe ansatz properties
are then exhibited. In section 5 we consider a non-compact vertex
model associated to the discrete $D_{\bf s}^{-}$ representation of the
$SL(2,{\cal R})$ symmetry. We present its algebraic Bethe ansatz
properties and argue that
they can be viewed as an appropriate limit of that  derived for
the vertex model defined in section \ref{sec41}.  Our conclusions are
presented in section 6. In Appendices A, B,  C and D
we summarize technical details that are useful for the
comprehension of the main text.  

\section{Definitions and Results}
\label{sec2}

We shall here review our previous general results for the algebraic Bethe
ansatz solution for the $U(1)$ invariant vertex models \cite{CM}. These
vertex models are statistical systems defined on a square lattice of
$L$ rows and $L$ columns whose intersections are denominated vertices. The
statistical configurations of these models
are characterized by assigning to
each lattice edge a variable that takes value on a set of integer numbers
$\{1,2,\cdots,N\}$. The Boltzmann weight at the $i$-th vertex is generally represented by
$R(\lambda,\mu_i)_{a,b}^{c,d}$ where $a,b,c,d=1, \cdots, N$ and the parameters $\mu_i$ play
the role of horizontal inhomogeneities.
The underlying $U(1)$ symmetry implies that the vertex weights 
satisfy the following ice rule constraint,  
\bear
\bullet~R(\lambda,\mu)_{a,b}^{c,d} \ne 0,& ~~~~ \mbox{for} ~~ a+b=c+d .
\nonumber \\
\bullet~R(\lambda,\mu)_{a,b}^{c,d}=0 ,& ~~~~ \mbox{for} ~~ a+b\ne c+d .
\label{ice}
\ear

As usual the integrability of the vertex model is guaranteed by imposing the 
Yang-Baxter equation for a 
$N^2 \times N^2$ $R$-matrix $R(\lambda,\mu)$ which in the notation of \cite{CM} is given
by, 
\EQ
R(\lambda,\mu)=\sum_{a,b,c,d=1}^{N} R(\lambda,\mu_i)_{a,b}^{c,d}  e_{a,c} \otimes e_{b,d},
\EN
where $ e_{a,b}$ denote $N\times N$ Weyl matrices.

The monodromy matrix associated to vertex models is build up by
considering the product of weights on the horizontal line which formally
can be written as, 
\EQ
\label{mono1}
{\cal T}_{\cal A}(\lambda) =
{\cal L}_{{\cal A}L}(\lambda,\mu_{L}) {\cal L}_{{\cal A}L-1}(\lambda,\mu_{L-1}) \dots
{\cal L}_{{\cal A}1}(\lambda,\mu_{1}).
\EN

The expression for the operators
${\cal L}_{{\cal A} i}(\lambda,\mu_i)$  in terms of the statistical weights is,
\EQ
\label{lope}
{\cal L}_{{\cal A} i}(\lambda,\mu_i)=\sum_{a,b,c,d=1}^{N} R(\lambda,\mu_i)_{a,b}^{c,d}  e_{a,c} \otimes e_{b,d}^{(i)} ,
\EN
where $e_{b,d}^{(i)}$ are $N\times N$ Weyl matrices acting 
on the quantum space
$\displaystyle{\prod_{i=1}^{L}\otimes _{i}^{N}}$  of a one-dimensional chain of length $L$. 

It turns out that the ice rule (\ref{ice}) permits us to construct a reference state in terms of
the tensor product of standard local ferromagnetic vectors $\ket{{\bf s}}_i$ with spin
${\bf s}=(N-1)/2$ for the monodromy matrix (\ref{mono1},\ref{lope}). This pseudo-vacuum state is,
\EQ
\ket{0}=\prod_{i=1}^{L} \otimes \ket{{\bf s}}_{i}, ~~~~ \ket{{\bf s}}_{i}=\left(\begin{array}{c}
1 \\ 0 \\ \vdots \\ 0 \end{array}\right)_{N},
\EN
where functions
$\omega_a(\lambda)$ introduced in Eq.(\ref{action}) are,
\EQ
w_{a}(\lambda)=\prod_{i=1}^{L} R(\lambda,\mu_{i})_{a, 1}^{a, 1} .
\label{omeg}
\EN

The other eigenstates of $T(\lambda)$ are constructed in terms of  a linear combination
of product of creation operators defined by 
the first row of the monodromy matrix. Due to the $U(1)$ symmetry these states are viewed as   
$n$-particle states 
parameterized by the number of rapidities $\lambda_1,\cdots,\lambda_n$ which can be
written as,
\EQ
\ket{\Phi_n}=\phi_n(\lambda_1,\cdots,\lambda_n) \ket{0}.
\EN

The states $\ket{\Phi_n}$  can be interpreted as 
excitations of spin $(N-1)/2-n$ over the reference state $\ket{0}$. The 
 mathematical structure of the vector
$\phi_n(\lambda_1,\cdots,\lambda_n) $ is given by the following $(N-1)$-step recurrence relation,
\bear
\phi_n(\lambda_1,\cdots,\lambda_n) &=& \sum_{e=1}^{m(n,N-1)}
{\cal T}_{1, 1+{e}}(\lambda_1) \sideset{}{^*} \sum_{\stackrel{2
\le j_2 <\dots <j_{e} \le n}{2 \le j_{e+1} < \dots
<j_{n} \le n}} \phi_{n-e}(\lambda_{j_{e+1}}, \dots,
\lambda_{j_{n}} )~
{_{e-1}}{\cal F}_{e-1}^{(2)} (\lambda_1,
\lambda_{j_2},\dots,\lambda_{j_{e}} )
\nonumber \\
&\times &
\prod_{k_1=2}^{e}
{\cal T}_{1,1}(\lambda_{j_{k_1}}) \prod_{ k_2=e+1 }^n
\frac{R(\lambda_{j_{k_2}},\lambda_{j_{k_1}})_{1,1}^{1,1}}{R(\lambda_{j_{k_2}},\lambda_{j_{k_1}})_{2,1}^{2,1}}
\theta_<(\lambda_{j_{k_2}},\lambda_{j_{k_1}}), 
\label{eing} 
\ear
where the symbol $*$ means that terms with $j_k=j_l$ are excluded in the sum and $m(x,y)$ denotes
the minimum integer of the pair $\{ x,y \}$.

The functions entering Eq.(\ref{eing}) are well defined in terms of the $R$-matrix elements
apart from the overall normalization  
${_{0}}{\cal F}_{0}^{(2)} (\lambda)$  which can be set to unity. In particular, 
$\theta_<(\lambda_{i},\lambda_{j})$ is defined by,
\EQ
\theta_<(\lambda_{i},\lambda_{j})
= \begin{cases} \displaystyle
\theta(\lambda_{i},\lambda_{j}),~~~~\mbox{for} ~~ i < j \cr
\displaystyle 1, ~~~~\mbox{for} ~~ i \ge j.
\label{theta<}
\end{cases}
\EN
where the expression for $\theta(\lambda,\mu)$ 
is, 
\EQ
\theta(\lambda, \mu)=
\frac{\left|
\begin{array}{cc}
R(\lambda,\mu)_{2,2}^{2,2} & R(\lambda,\mu)_{3,1}^{2,2} \\
R(\lambda,\mu)_{2,2}^{3,1} & R(\lambda,\mu)_{3,1}^{3,1}
\end{array} \right|}{R(\lambda,\mu)_{1,1}^{1,1}
R(\lambda,\mu)_{3,1}^{3,1}}.
\label{theta}
\EN

The auxiliary function
${_{b}}{\cal F}_{b}^{(2)} (\lambda_1,
\lambda_{2},\dots,\lambda_{b+1})$ is a special class of more general
functions 
that are companion of the unwanted
vectors generated by the action of the monodromy matrix elements 
${\cal T}_{a,a}(\lambda)$ on the $b$-particle state $\ket{\Phi_{b}}$.   
We shall denote these generalized 
off-shell amplitudes by ${_{c}}{\cal F}_{b}^{(a)} (\lambda_1,
\lambda_{2},\dots,\lambda_{b+1})$ 
where the range of the extra indices are 
$c=0,\cdots,b$ and $a=1,\cdots,N-b$. Here we remark that the index $c$ accounts
for the number of weights $\omega_1(\lambda_i)$ that is present in the respective
undesirable term proportional to the operators 
$\displaystyle{\prod_{i=1}^{c} \omega_1(\lambda_i) 
{\cal T}_{a,a+b}(\lambda)}$.   
It turns out that they satisfy a set of recurrence relations
whose initial conditions are, 
\EQ
\label{psi1}
{_0}{\cal F}_{1}^{(a)}(\lambda,\mu) =
-{_1}{\cal F}_{1}^{(a)}(\lambda,\mu) =
\frac{R(\lambda,\mu)_{a+1, 1}^{a, 2}}{R(\lambda,\mu)_{a+1, 1}^{ a+1, 1}},~~\mathrm{for}~~a=1,\cdots,N-1.
\EN

The structure of the off-shell amplitudes for
$c \neq 0$ and $c \neq b$ has a direct factorized form of
the following type,
\bear {_c}{\cal
F}_{b}^{(a)}(\lambda,\lambda_1,\dots,\lambda_b) &=& {_0}{\cal
F}_{b-c}^{(a)}(\lambda,\lambda_{(c+1)},\dots,\lambda_{b}) {_c}{\cal
F}_{c}^{(a+b-c)}(\lambda,\lambda_1,\dots,\lambda_c) \prod_{i=c+1}^b
\prod_{j=1}^c \frac{R(\lambda_i,\lambda_j)_{1, 1}^{1, 1}}
{R(\lambda_i,\lambda_j)_{2, 1}^{2, 1}} \nonumber \\
&& \mbox{for}~~b=2,\dots,N-1; ~a=1,\dots,N-b;~c=1,\dots,b-1.
\label{fbc}
\ear

However, in order to iterate Eq.(\ref{fbc}) we still need to know the
off-shell amplitudes for the  extremum values $c=0$ and $c=b$. These functions
satisfy more complicated recurrence relations
involving the sum of products of many distinct terms, namely
\bear {_0}{\cal
F}_{b}^{(a)}(\lambda,\lambda_1,\dots,\lambda_b) &=&
\sum_{e=1}^b \frac{R(\lambda,\lambda_1)_{a+e, 1}^{a,
1+e}}{R(\lambda,\lambda_1)_{a+b, 1}^{a+b, 1}} \sideset{}{^*}
\sum_{\stackrel{2 \le j_1< \dots < j_{(b-{e})} \le b}{2 \le
j_{(b-{e}+1)}< \dots < j_{(b-1)} \le b}} {_0}{\cal
F}_{b-{e}}^{(a+{e})}(\lambda,\lambda_{j_1},\dots,
\lambda_{j_{(b-{e})}}) \nonumber \\
& \times & {_{{e}-1}}{\cal
F}_{{e}-1}^{(2)}(\lambda_1,\lambda_{j_{(b-{e}+1)}},\dots,\lambda_{j_{(b-1)}})
\prod_{l_1=1}^{b-{e}} \prod_{l_2=b-{e}+1}^{b-1}
\frac{R(\lambda_{j_{l_1}},\lambda_{j_{l_2}})_{1, 1}^{1, 1}}
{R(\lambda_{j_{l_1}},\lambda_{j_{l_2}})_{2, 1}^{2, 1}}
\theta_<(\lambda_{j_{l_1}},\lambda_{j_{l_2}}) \nonumber \\
&\mbox{for}& ~~~~~~~~~~~~~ b=1,\dots,N-1; ~~a=1,\dots,N-b,
\label{fb0}
\ear
and
\bear {_b}{\cal
F}_{b}^{(a)}(\lambda,\lambda_1,\dots,\lambda_b) &=& -
\sum_{{f}=0}^{b-1} ~\sum_{1 \le l_1 < l_2 < \dots <
l_{(b-{f})} \le b} {_{{f}}}{\cal F}_{b}^{(a)}(\lambda,\{
\lambda_i \}_{i \ne  l_1,\dots,l_{(b-{f}) }}^{i=1,\dots,b} ,
\lambda_{l_1},\dots,\lambda_{l_{(b-{f})}}) \nonumber \\
&\times & \prod_{s=1}^{b-{f}} \prod_{\stackrel{i=1}{i \ne
l_1,\dots,l_{(b-{f})} }}^{b}
\theta_<(\lambda_{i},\lambda_{l_s})
\frac{R(\lambda_{i},\lambda_{l_s})_{1, 1}^{1,
1}}{R(\lambda_{i},\lambda_{l_s})_{2, 1}^{2, 1}}
\frac{R(\lambda_{l_s},\lambda_{i})_{2, 1}^{2,
1}}{R(\lambda_{l_s},\lambda_{i})_{1, 1}^{1, 1}}
\nonumber \\
&\mbox{for}& ~~~~~~~~~~~~~ b=1,\dots,N-1; ~~a=1,\dots,N-b,
\label{fbb}
\ear
where 
$\{
\lambda_i \}_{i \ne  l_1,\dots,l_{p} }^{i=1,\dots,b}$  means that out of
the possible variables $\lambda_1,\cdots,\lambda_b$ those indexed by
$\lambda_{l_1},\cdots,\lambda_{l_p}$ are absent in the set.

At this point we emphasize that relations (\ref{psi1}-\ref{fbb}) provide a self-consistent
way to determine all the off-shell amplitudes entering in a given $n$-particle state. For example,
in order to generate the two-particle off-shell amplitudes one has to substitute the
one-particle initial condition (\ref{psi1}) 
in expressions (\ref{fbc}-\ref{fbb}). These data together are then able to provide us the
three-particle off-shell amplitudes and this procedure can then be iterated  
until we reach the final step with the total number of $(N-1)$ particles. We observe that in practice,
for the case of a specific model, it is sufficient to find closed expressions for the
amplitudes
${_0}{\cal
F}_{b}^{(a)}(\lambda,\lambda_1,\dots,\lambda_b)$  and
${_b}{\cal
F}_{b}^{(a)}(\lambda,\lambda_1,\dots,\lambda_b)$ since the remaining ones 
are automatically fixed by
Eq.(\ref{fbc}).  

We now turn our attention to the action of the transfer matrix $T(\lambda)$
on the multi-particle state $\ket{\Phi_n}$. Following \cite{CM} we find that 
$T(\lambda) \ket{\Phi_n}$ can be rewritten as,
\bear 
T(\lambda) \ket{\Phi_n} &=&
\sum_{a=1}^N w_a(\lambda) \prod_{i=1}^{n}P_{a}(\lambda,\lambda_{i}) \ket{\Phi_n}
- \sum_{t=1}^n \sum_{a=1}^{N-t} {\cal T}_{a, a+t}(\lambda)
\sum_{p=0}^{t-1} \sideset{}{^*}\sum_{\stackrel{1 \le j_1 < \dots <
j_{p} \le n}{1 \le j_{(p+1)} <
 \dots < j_{t} \le n}}
\nonumber \\
& \times &
{_p}{\cal 
F}_{t}^{(a)}(\lambda,\lambda_{j_1},\dots,\lambda_{j_t}) 
{_p}{\cal H}_t^{(n)}(
\lambda_{j_1},\dots,\lambda_{j_t}) 
\phi_{n-t}(\{ \lambda_{i} \}_{i \ne
j_1,\dots,j_{t} }^{i=1,\dots,n})\ket{0},
\label{tphin}
\ear
where functions 
$P_{a}(\lambda,\mu)$  proportional to the states $\ket{\Phi_n}$ are, 
\EQ
P_{a}(\lambda,\mu) = \begin{cases}
\displaystyle
\frac{R(\mu,\lambda)_{1,1}^{1,1}}{R(\mu,\lambda)_{2,1}^{2,1}},~~~~\mbox{for} ~~ a=1 \cr
\displaystyle
\frac{\left| \begin{array}{cc}
R(\lambda,\mu)_{a,2}^{a,2} & R(\lambda,\mu)_{a+1,1}^{a,2} \\
R(\lambda,\mu)_{a,2}^{a+1,1} & R(\lambda,\mu)_{a+1, 1}^{a+1, 1}
\end{array} \right|}{R(\lambda,\mu)_{a, 1}^{a, 1}R(\lambda,\mu)_{a+1,1}^{a+1,1}}, ~~~~ \mbox{for} ~~ 2 \le a \le N-1 \cr
\displaystyle
\frac{R(\lambda,\mu)_{N,2}^{N,2}}{R(\lambda,\mu)_{N,1}^{N,1}}, ~~~~\mbox{for} ~~ a=N.
\label{pa}
\end{cases}
\EN

The expression (\ref{tphin}) tells us that the unwanted terms, 
i.e the vectors not parallel to the state $\ket{\Phi_n}$, have 
a universal pattern governed by the following type of operators
${\cal T}_{a, a+t}(\lambda)
\phi_{n-t}(\{ \lambda_{i} \}_{i \ne
j_1,\dots,j_{t} }^{i=1,\dots,n})$ for $t=1, \cdots, m(n,N-1)$. We also see that the respective amplitudes are 
product of two distinct classes of functions  
${_p}{\cal 
F}_{t}^{(a)}(\lambda,\lambda_{j_1},\dots,\lambda_{j_t})$ and 
${_p}{\cal H}_t^{(n)}(
\lambda_{j_1},\dots,\lambda_{j_t})$. The first part carries a dependence on both the spectral parameter $\lambda$ and
the Bethe rapidities $\lambda_1,\cdots,\lambda_n$ 
and satisfies the non-trivial recurrence relations (\ref{psi1}-\ref{fbb}). By contrast, the second one depends only on
the Bethe variables and has a rather simple factorized form,
\bear 
{_p}{\cal H}_t^{(n)}(
\lambda_{j_1},\dots,\lambda_{j_t}) &=&
\prod_{s=1}^{p} w_1(\lambda_{j_s})
\prod_{\stackrel{i=1}{i \ne  j_1,\dots,j_{t} }}^{n}
\frac{R(\lambda_{i},\lambda_{j_s})_{1, 1}^{1,
1}}{R(\lambda_{i},\lambda_{j_s})_{2, 1}^{2, 1}}
\theta_<(\lambda_{i},\lambda_{j_s}) 
\prod_{r=p+1}^{t} \prod_{s=1}^{p}
\theta_<(\lambda_{j_s},\lambda_{j_r}) 
\nonumber \\
& \times &
\prod_{\stackrel{i=1}{i \ne
j_1,\dots,j_{t} }}^{n} 
\theta_<(\lambda_{i},\lambda_{j_r})
%\nonumber \\
%& \times &
 \left[ \prod_{r=p+1}^{t} w_2(\lambda_{j_r})
\prod_{\stackrel{i=1}{i \ne j_1,\dots,j_{t} }}^{n}
\frac{R(\lambda_{j_r},\lambda_{i})_{1, 1}^{1,
1}}{R(\lambda_{j_r},\lambda_{i})_{2, 1}^{2, 1}}
\theta(\lambda_{j_r},\lambda_{i}) 
\prod_{s=1}^{p} \theta(\lambda_{j_r},\lambda_{j_s}) \right.
\nonumber \\
&-& \left. \prod_{r=p+1}^{t} w_1(\lambda_{j_r}) \prod_{\stackrel{i=1}{i \ne
j_1,\dots,j_{t} }}^{n} \frac{R(\lambda_{i},\lambda_{j_r})_{1, 1}^{1,
1}}{R(\lambda_{i},\lambda_{j_r})_{2, 1}^{2, 1}} \prod_{s=1}^{p}
\frac{R(\lambda_{j_s},\lambda_{j_r})_{1, 1}^{1,
1}}{R(\lambda_{j_s},\lambda_{j_r})_{2, 1}^{2, 1}}
\frac{R(\lambda_{j_r},\lambda_{j_s})_{2, 1}^{2,
1}}{R(\lambda_{j_r},\lambda_{j_s})_{1, 1}^{1, 1}} \right],
\nonumber \\
\label{hhh} 
\ear
which is 
easily computed from the knowledge of few statistical weights. In fact it only depends
on the amplitudes $R(\lambda,\mu)_{a,1}^{a,1}$ and function $\theta(\lambda,\mu)$.

In order to enforce that $\ket{\Phi_n}$ is an eigenstate of $T(\lambda)$
we need to find variables $\lambda_1,\cdots,\lambda_n$ such that  all the
unwanted terms vanish for arbitrary values of the spectral parameter. Considering
the functional form of functions proportional
to the undesirable terms we conclude
that this is achieved by imposing that
${_p}{\cal H}_t^{(n)}(
\lambda_{j_1},\dots,\lambda_{j_t})=0$. From Eq.(\ref{hhh}) it is not difficult
to see that this leads us to the following Bethe
equations for the rapidities $\lambda_1,\cdots,\lambda_n$,
\EQ
\frac{w_1(\lambda_{j})}{w_2(\lambda_{j})} =
\prod_{\stackrel{i=1}{i \ne j}}^{n}
\theta(\lambda_{j},\lambda_{i})
\frac{R(\lambda_{j},\lambda_{i})_{1, 1}^{1,
1}}{R(\lambda_{j},\lambda_{i})_{2, 1}^{2, 1}}
\frac{R(\lambda_{i},\lambda_{j})_{2, 1}^{2,
1}}{R(\lambda_{i},\lambda_{j})_{1, 1}^{1, 1}}~~~~
\mbox{for}~~j=1,\dots,n. \label{beaN}
\EN

As a consequence of that we conclude that the
$n$-particle transfer matrix 
eigenvalue associated to the on-shell Bethe states,
\EQ
T(\lambda) \ket{\Phi_n} = \Lambda_n(\lambda) \ket{\Phi_n},
\EN
is determined by the expression,
\EQ \Lambda_{n}(\lambda) =
\sum_{a=1}^N w_a(\lambda) \prod_{i=1}^{n}P_{a}(\lambda,\lambda_{i}).
\label{gaman}
\EN

In next sections we shall present the explicit expressions for the non-trivial
off-shell amplitudes (\ref{psi1}-\ref{fbb}), the Bethe ansatz equations (\ref{beaN})
and the eigenvalues (\ref{gaman}) in the case of three different types of vertex models.

\section{The XXZ-${\bf s}$ model}

The study of integrable higher spin Heisenberg XXZ-${\bf s}$ chains started with 
the search of $N$-state vertex model solutions of 
Yang-Baxter equation for 
$N=3,4$ \cite{FA1,SO}. 
The respective
$R$-matrix for arbitrary values of ${\bf s}=(N-1)/2$ was first proposed within  
the fusion procedure 
in the special case of the isotropic $SU(2)$ XXX-${\bf s}$ chain \cite{RKS}.
Other progress on higher spin descendants of the Heisenberg 
model has been made with the notion of universal $R$-matrix \cite{FAT}, 
the connection 
with higher spin representations of the 
$U_q[SU(2)]$ algebra for generic values of $q$ \cite{JI} and 
the relationship with new link polynomials  
and generalized braid monoid algebras \cite{WA1}. 
The $R$-matrix associated to the XXZ-${\bf s}$ chain 
is conveniently written with 
the help of an auxiliary matrix $\check{R}(\lambda)$,
\EQ
R(\lambda,\mu)=P\check{R}(\lambda,\mu),
\label{rmatrix}
\EN
where $P=\sum_{a,b=1}^N e_{a,b} \otimes e_{b,a}$ is the $C^N \otimes C^N$ permutator.

The matrix $\check{R}(\lambda,\mu)$ can be expressed in a closed form by using the projectors 
${\check P}_j(q)$ on the tensor product of two irreducible 
representations of $U_q[SU(2)]$ with spin ${\bf s}=(N-1)/2$ \cite{JI},
\EQ
\check{R}(\lambda)=\sum_{j=0}^{N-1}
\prod_{k=1}^{j}
\frac{\sinh[ \imath k \gamma +\lambda-\mu]}{\sinh[\imath k \gamma -\lambda+\mu]}
{\check P}_j(q),
\EN
where the deformation parameter $q$ and the the anisotropy $\gamma$ are related by
$q=\exp[-2 \gamma \imath]$.

The expressions for the operators ${\check P}_j(q)$ can in principle be given by means of an 
interpolation among the roots of the $U_q[SU(2)]$ Casimir operator \cite{BI}. Though ${\check P}_j(q)$
can be expressed in terms of such operator in a simple way its explicit expression
on the Weyl basis requires an extra amount of work. This step
is essential to provide us the weights $R_{a,b}^{c,d}(\lambda,\mu)$ which
are the main ingredient to establish a statistical mechanics interpretation and
to compute the respective Bethe ansatz properties as described in section 2.
We found that these projectors can be written as,
\EQ
{\check P}_j(q)
=
\sum_{\stackrel{k=0}{k \ne j}}^{N-1}
\frac{\hat{S}(q)-(-1)^k q^{k (k+1)/2}}
{(-1)^j q^{j (j+1)/2}- (-1)^k q^{k (k+1)/2}}.
\label{proj}
\EN

The operator $\hat{S}(q)$ in Eq.(\ref{proj}) play the role 
of a braid satisfying
a form of the Yang-Baxter equation without spectral parameter, namely
\EQ
[\hat{S}(q) \otimes I_N][I_N \otimes \hat{S}(q)][\hat{S}(q) \otimes I_N]=[I_N \otimes \hat{S}(q)][\hat{S}(q) \otimes I_N][I_N 
\otimes \hat{S}(q)],
\label{braidrel}
\EN
where $I_N$ is the $N \times N$ identity matrix.

In Weyl basis the expression for the braid $\hat{S}(q)$ is,
\EQ
\hat{S}(q)= \sum_{\stackrel{a,b,c,d=1}{a \ge d; c \ge b }}^{N}
S_{c,d}^{a,b}(q)
e_{b,d}
\otimes
e_{a,c},
\EN
where the weights  $S_{c,d}^{a,b}(q)$ are given by
\bear
S_{c,d}^{a,b}(q)
=-(-1)^{N}
\frac{q^{\left[\frac{N(N-1)}{2}+\frac{(b-1)(d-N)}{2} +\frac{(d-1)(b-N)}{2} \right] }}
{\sqrt{W_0(a-d) W_0(c-b)}}
\sqrt{
\prod_{\epsilon=0}^{1}
\frac{ W_{\epsilon}(a-1) W_{\epsilon}(c-1)}
{ W_{\epsilon}(d-1) W_{\epsilon}(b-1)}
} \delta_{a+b,c+d},
\label{braidXXZ}
\ear
while 
function $W_{\epsilon}(n)$ for $\epsilon =0,1$ is defined by the following product,
\EQ
W_{\epsilon}(n)
=
\prod_{k=1}^{n}
(1-q^{k-\epsilon N}).
\label{wfunction}
\EN

The relevant feature of Eqs.(\ref{rmatrix}-\ref{wfunction}) is that they can be easily implemented to 
compute the $R$-matrix for relatively large values of $N$. By substituting the $R$-matrix amplitudes
in Eqs.(\ref{beaN},\ref{gaman}) we are able to obtain the respective on-shell
Bethe ansatz properties. After some manipulations we find that the transfer matrix
eigenvalues are,
\bear
\Lambda_n(\lambda) &=&
\sum_{a=1}^{N}
\prod_{l=1}^{L}
\prod_{k=1}^{a-1}
\frac{\sinh[\lambda-\mu_l-\imath (k-1) \gamma]}{\sinh[\lambda-\mu_l+\imath (N-k) \gamma]}
\nonumber \\
& \times & \prod_{i=1}^{n}
\frac{\sinh[\lambda-\lambda_i-\imath (N-1)\gamma] \sinh[\lambda-\lambda_i+\imath \gamma]}
{\sinh[\lambda-\lambda_i-\imath (a-1)\gamma] \sinh[\lambda-\lambda_i-\imath (a-2)\gamma]},
\label{eigXXZ}
\ear
while the Bethe ansatz equations for the variables $\lambda_j$ are\footnote{The Bethe ansatz
equations (\ref{beaXXZ}) can be made more 
symmetrical by performing the shift $\lambda_j \rightarrow
\lambda_j -\imath \frac{(N-1)\gamma}{2} $.} 
\EQ
\prod_{l=1}^{L}
\frac{\sinh[\lambda_j-\mu_l+\imath (N-1) \gamma]}{\sinh[\lambda_j-\mu_l]}
 =
\prod_{\stackrel{i=1}{i \neq j}}^{n}
\frac{\sinh[\lambda_j-\lambda_i+\imath \gamma ]}{\sinh[\lambda_j-\lambda_i-\imath \gamma]}
,~~~j = 1, \dots, n. 
\label{beaXXZ}
\EN

We remark that the Bethe ansatz results (\ref{eigXXZ},\ref{beaXXZ}) have been obtained 
before by the mechanism of fusion \cite{BAB1,RK}.
This method allows us to write the transfer matrix of higher spin systems 
by means of traces taken on the smaller two-dimensional ${\bf s}=1/2$
auxiliary space and the respective eigenvalues are determined recursively.
The Bethe equations are then proposed by requiring the analyticity of the eigenvalues 
rather than by explicit cancellation of the unwanted terms preventing us information on
the off-shell data. 
By contrast, the results described in section 2 provide us the means to determine
the off-shell Bethe ansatz structure by working out the 
recurrence relations (\ref{psi1}-\ref{fbb}). This step,
however, requires a considerable amount of additional work specially when we are 
interested on the results for arbitrary
values of $N$. In appendix A we summarize the technical details that we have
devised to perform this computation for all the models considered in this
paper. By implementing the analysis of Appendix A for the XXZ-${\bf s}$ model
we find that 
the expressions for the basic amplitudes 
${_0}{\cal F}_b^{(a)}(\lambda,\lambda_1,\dots,\lambda_b)$ and
${_b}{\cal F}_b^{(a)}(\lambda,\lambda_1,\dots,\lambda_b)$  factorize 
in terms of elementary trigonometric 
functions,
\EQ
{_0}{\cal F}_1^{(a)}(\lambda,\mu)
=
-{_1}{\cal F}_1^{(a)}(\lambda,\mu)
=\exp[\mu-\lambda]
\frac{
\sqrt{
\sinh \left[\imath (N-1)\gamma \right]
\sinh \left[\imath (N-a)\gamma \right]
\sinh \left[\imath a \gamma \right]
}}
{
\sqrt{
\sinh \left[\imath \gamma \right]
}
\sinh \left[\imath (a-1)\gamma -\lambda+\mu \right]
},
\EN
\bear
{_0}{\cal F}_b^{(a)}(\lambda,\lambda_1,\dots,\lambda_b)
&=&
{\cal G}_{0}^{(a,b)}(\gamma)
\prod_{\stackrel{i,j=1}{i<j}}^{b}
\frac{\sinh \left[ \lambda_i-\lambda_j -\imath (N-1)\gamma\right]}{\sinh \left[ \lambda_i-\lambda_j-\imath \gamma \right]}
\prod_{i=1}^{b}
{_0}{\cal F}_1^{(a+b-1)}(\lambda,\lambda_i),
\\
{_b}{\cal F}_b^{(a)}(\lambda,\lambda_1,\cdots,\lambda_b)
&=&
{\cal G}_{0}^{(N+1-a-b,b)}(\gamma)
\prod_{\stackrel{i,j=1}{i<j}}^{b}
\frac{\sinh \left[ \lambda_i-\lambda_j -\imath (N-1)\gamma \right]}{\sinh \left[ \lambda_i-\lambda_j-\imath \gamma \right]}
\prod_{i=1}^{b}
{_1}{\cal F}_1^{(a)}(\lambda,\lambda_i).
\ear

The  constant
${\cal G}_{0}^{(a,b)}(\gamma)$ does not depend 
either on the spectral parameter $\lambda$ or on the 
variables $\lambda_1,\cdots,\lambda_n$ and it is given by,
\bear
{\cal G}_{0}^{(a,b)}(\gamma)
&=&
\prod_{l=1}^{b-1}
\sqrt{
\frac{\sinh \left[\imath (a+b-1-l)\gamma \right] }
{\sinh \left[\imath (a+b-1)\gamma \right] }
\frac{\sinh \left[\imath (N+1-a-l) \gamma \right] }
{\sinh \left[\imath (N+1-a-b)\gamma \right] }
}.
\ear

To complete the off-shell data we are only left with the computation of function
${_p}{\cal H}_t^{(n)}(
\lambda_{j_1},\dots,\lambda_{j_t})$ which depends on the weights $R(\lambda,\mu)_{a,1}^{a,1}$ and
the Bethe ansatz function $\theta(\lambda,\mu)$. For the XXZ-${\bf s}$ model they are given by,
\EQ
R(\lambda,\mu)_{a,1}^{a,1}=\prod_{k=1}^{a-1}
\frac{\sinh[\lambda-\mu-\imath (k-1) \gamma]}{\sinh[\lambda-\mu+\imath (N-k) \gamma]},
\EN
and 
\EQ
\theta(\lambda,\mu)=
\frac{\sinh[\lambda-\mu-\imath (N-1) \gamma]}{\sinh[\lambda-\mu+\imath (N-1) \gamma]}
\frac{\sinh[\lambda-\mu+\imath \gamma]}{\sinh[\lambda-\mu-\imath \gamma]}.
\label{thetaXXZ}
\EN

We close this section remarking that the off-shell part produced by a complete algebraic Bethe 
ansatz analysis contains much more information than we would think at first sight \cite{BAB}.
Indeed, the semi-classical limit of the off-shell terms play an important role in the 
solution of Gaudin like models and Knizhnik-Zamolodchikov equations \cite{BAB}.
Due to the factorizability of the off-shell data  
in terms of simple ``two-body'' functions their semi-classical study should not be complicated.
It is expected that such analysis will give us a representation for the solution of the trigonometric Knizhnik-Zamolodchikov 
equation corresponding to higher spin representations of $U_q[SU(2)]$. Recall that this study has so far
been pursued in the particular case of the six (${\bf s}=1/2$) and nineteen (${\bf s}=1$) vertex models \cite{BAB4,LIM}.

\section{Colored Vertex Models}

The aim of this section is to discuss the transfer matrix diagonalization of solvable
$U(1)$ vertex models directly connected to non-generic braid group representations first
discovered by Couture, Lee and Schmeing \cite{COT}. Such braid solution was then generalized to
include color variables on the braid strings leading to the proposition of new
$N$-state vertex models \cite{DEG,DEG3,DEG1}. These models have been also viewed as
Yang-Baxter solutions associated with the finite dimensional representation
of $U_q[SU(2)]$ when $q$ is a root of unity \cite{COT1,SIE}.

\subsection{Additive $R$-matrices}
\label{sec41}

From colored braid matrices it is possible to construct $R$-matrices that are additive
with respect to the spectral parameters. This happens when we consider that the
color variable attached to the $i$-th string is the same for all $i$-th indices
playing the role of an extra continuous parameter denoted here by $\bar{\gamma}$.
This variable characterizes  
the additional freedom of non-cyclic irreducible representation of
quantum group at roots of unity
\cite{COT1}. To avoid confusion with the deformation parameter of the
previous section we shall denote the roots of unity by the variable
$\omega$,
\EQ
\omega=\exp[\frac{2 \pi \imath k}{N}]~~\mathrm{for~ k~ and~ N~ coprime}.
\EN

This means that out of the possible values $k=1,\cdots, N-1$ 
the only admissible ones are those
that are prime with $N$.
The underlying $U(1)$ symmetry guarantees that the corresponding braid
$\hat{S}(\bar{\gamma},\omega)$ can once again be represented  as,
\EQ
\hat{S}(\bar{\gamma},\omega)= \sum_{\stackrel{a,b,c,d=1}{a \ge d; c \ge b }}^{N}
S_{c,d}^{a,b}(\bar{\gamma},\omega)
e_{b,d}
\otimes
e_{a,c}.
\label{bracol}
\EN

Considering the results of \cite{DEG} we find that the amplitudes 
$S_{c,d}^{a,b}(\bar{\gamma},\omega)$ can be written as,
\bear
S_{c,d}^{a,b}(\bar{\gamma},\omega) & =& \frac{\omega^{(b-1)(d-1)} \exp[\bar{\gamma}(b+d-2)]}
{H(\omega,a-d) }
\sqrt{
\frac{ H(\omega,a-1) H(\omega,c-1)}
{H(\omega,d-1) H(\omega,b-1)}}
\nonumber \\ 
& \times &
\sqrt{
\frac{H(\exp(2\bar{\gamma}),a-1) H(\exp(2\bar{\gamma}),c-1)}
{H(\exp(2\bar{\gamma}),d-1) H(\exp(2\bar{\gamma}),b-1)}}
\delta_{a+b,c+d},
\label{bracol1}
\ear
where function 
$H(x,n)$ denotes the following factorial product,
\EQ
H(x,n)=
\prod_{k=0}^{n-1}
(1-xw^k).
\label{Hfunction}
\EN

The construction of spectral parameter dependent $R$-matrices for models based on the  colored braids
was first discussed for general $N$ by Deguchi and Akutsu
\cite{DEG3,DEG1} within the quantum group framework.
Here we shall present an alternative manner to generate a solution of the Yang-Baxter equation from 
the braid representation (\ref{bracol}-\ref{Hfunction}).
This procedure is usually called Baxterization \cite{JO} and it is able to produce additive $R$-matrices directly on the
Weyl basis for arbitrary values of $N$. This analysis offers us a  practical computational way
to determine the Boltzmann weights and to our knowledge it is original in the literature. 
The first step in this method is to examine
the eigenvalue structure of the braid \cite{CH}. The diagonalization of the braid (\ref{bracol},\ref{bracol1}) reveals us
that it satisfies the following polynomial relation,
\EQ
\prod_{i=1}^{N}[\hat{S}(\bar{\gamma},\omega)-\xi_i I_N \otimes I_N]=0,
\EN
where the $N$ distinct eigenvalues $\xi_i$ are,
\EQ
\xi_i=(-1)^{i+1} w^{\frac{(i-2)(i-1)}{2}} \exp[2\bar{\gamma}(i-1)].
\EN

The knowledge of the eigenvalues of the braid permits us to formally decompose it as 
$\hat{S}(\bar{\gamma},\omega)=\displaystyle{\sum_{i=1}^N \xi_i \check{P}_i(\bar{\gamma},\omega)}$ where
$\check{P}_i(\bar{\gamma},\omega)$ is the projector on the subspace $\xi_i$,
\EQ
\check{P}_i(\bar{\gamma},\omega)=\prod_{\stackrel{k=1}{k \ne i}}^N 
\frac{\hat{S}(\bar{\gamma},\omega)-\xi_k}{\xi_i-\xi_k}.
\EN

The form of the $R$-matrices derived in the context of the quantum group framework 
suggests us to consider the following ansatz for $\check{R}(\lambda)$,
\EQ
\check{R}(\lambda,\mu)=\sum_{i=1}^N \rho_i(\lambda-\mu) \check{P}_i(\bar{\gamma},\omega).
\label{rcol}
\EN

The scalar functions $\rho_i(\lambda)$ can be fixed by means of the unitarity property 
$\check{R}(\lambda,\mu) 
\check{R}(\mu,\lambda)= I_N \otimes I_N$ 
as well as by imposing that the original braid should be recovered when one
takes the spectral parameter to infinity. The simplest functional form for 
$\rho_i(\lambda)$ fulfilling such properties is,
\EQ
\rho_i(\lambda)=
\prod_{k=i}^{N-1}
\frac{\left(1+\exp\left[2 \lambda \right] \frac{\xi_{k+1}}{\xi_k}\right)}{
\left(\exp\left[2 \lambda \right] + \frac{\xi_{k+1}}{\xi_k}\right)}.
\label{rhocol}
\EN

Now the remaining freedom we have at hand to fix the underlying $R$-matrix
is only concerned with the permutation of the $N$ eigenvalues $\xi_i$. The
suitable ordering of $\xi_i$ is selected out by imposing that the ansatz (\ref{rcol},\ref{rhocol}) for the
$R$-matrix indeed satisfy the Yang-Baxter equation. Putting the above considerations altogether we find
that the corresponding $R$-matrix is,
\EQ
\check{R}(\lambda,\mu)=\sum_{l=1}^{N}
\prod_{j=l}^{N-1}
\frac{\sinh[ \frac{\imath \pi k(j-1)}{N}+\bar{\gamma}+\lambda-\mu]}{\sinh[\frac{\imath \pi  k(j-1)}{N}+\bar{\gamma}-\lambda+\mu]}
P_l(\bar{\gamma},\omega).
\label{rcols}
\EN

As before the $R$-matrix representation (\ref{rcols}) provides us the means to compute
the respective weights $R_{a,b}^{c,d}(\lambda,\mu)$ for moderate large
values of $N$. This is the basic ingredient to determine the Bethe ansatz
properties of this type of vertex model. Once again by using Eqs.(\ref{beaN},\ref{gaman})
we find that the corresponding eigenvalue is,
\bear
\Lambda_n(\lambda) &=&
\sum_{a=1}^{N}
\prod_{l=1}^{L}
\prod_{j=1}^{a-1}
\frac{\sinh[\lambda-\mu_l+\frac{\imath \pi k}{N}(j-1) ]}{\sinh[\lambda-\mu_l+\bar{\gamma}+\frac{\imath \pi k}{N}(j-1)]}
\nonumber \\
& \times & \prod_{i=1}^{n}
\frac{\sinh[\lambda-\lambda_i-\bar{\gamma}] \sinh[\lambda-\lambda_i-\frac{\imath \pi k}{N}]}
{\sinh[\lambda-\lambda_i-\frac{\imath \pi k}{N}(1-a)] \sinh[\lambda-\lambda_i-\frac{\imath \pi k}{N}(2-a)]},
\label{eigcol}
\ear
provided that the rapidities $\lambda_j$ satisfy the following
Bethe ansatz equations,
\EQ
\prod_{l=1}^{L}
\frac{\sinh[\lambda_j-\mu_l+\bar{\gamma}]}{\sinh[\lambda_j-\mu_l]}
 =
\prod_{\stackrel{i=1}{i \neq j}}^{n}
\frac{\sinh[\lambda_j-\lambda_i-\frac{\imath \pi k}{N}]}{\sinh[\lambda_j-\lambda_i+\frac{\imath \pi k}{N}]}
,~~~j = 1, \dots, n. 
\label{beacol}
\EN

Here we remark that the on-shell Bethe ansatz results (\ref{eigcol},\ref{beacol})
have been previously discussed in the literature \cite{SIE1}. The derivations are sketched according
to the lines used to solve the XXZ-${\bf s}$ chain through a partial algebraic Bethe
ansatz analysis. The Bethe
ansatz equations are obtained by using the hypothesis of analyticity of the proposed eigenvalues
and the off-shell structure is not presented. However, we point out  
that even the on-shell Bethe ansatz
results proposed in \cite{SIE1} are not complete. The main 
branch $k=1$ for $N$ odd is not predicted as well as the many other possible
choices of $k$ for a given $N$ have been overlooked. There exists also an
unnecessary distinction whether the dimension of the
representation is even or odd. In this sense our findings (\ref{eigcol},\ref{beacol})
provide a non-trivial complement to those proposed in \cite{SIE1} for the on-shell
properties.

We shall now discuss the results for the corresponding off-shell amplitudes. This is
again done along the lines described in
Appendix A. However, the computations of the respective constants are more
cumbersome than that
of the XXZ-${\bf s}$ model due to the many
possible branches $k$ for each $N$. 
The final results are,
\EQ
{_0}{\cal F}_1^{(a)}(\lambda,\mu)=
-{_1}{\cal F}_1^{(a)}(\lambda,\mu)
=
\exp[\mu-\lambda]
\frac{
\sqrt{
\sinh \left[\bar{\gamma} \right]
\sinh \left[\frac{\imath \pi k}{N} a \right]
\sinh \left[\bar{\gamma} + \frac{\imath \pi k}{N} (a-1) \right]
}}
{
\sqrt{
\sinh \left[\frac{\imath \pi k}{N} \right]
}
\sinh \left[\mu-\lambda- \frac{\imath \pi k}{N} (a-1) \right]
},
\EN
\bear
{_0}{\cal F}_b^{(a)}(\lambda,\lambda_1,\dots,\lambda_b)
&=&
{\cal G}_{1}^{(a,b)}(\bar{\gamma})
\prod_{\stackrel{i,j=1}{i<j}}^{b}
\frac{\sinh \left[ \lambda_i-\lambda_j -\bar{\gamma}\right]}{\sinh \left[ \lambda_i-\lambda_j + \frac{\imath \pi k}{N} \right]}
\prod_{i=1}^{b}
{_0}{\cal F}_1^{(a+b-1)}(\lambda,\lambda_i),
%\nonumber \\
%&&
\\
{_b}{\cal F}_b^{(a)}(\lambda,\lambda_1,\dots,\lambda_b)
&=&
{\cal G}_{2}^{(a,b)}(\bar{\gamma})
\prod_{\stackrel{i,j=1}{i<j}}^{b}
\frac{\sinh \left[ \lambda_i-\lambda_j -\bar{\gamma}\right]}{\sinh \left[ \lambda_i-\lambda_j + \frac{\imath \pi k}{N} \right]}
\prod_{i=1}^{b}
{_1}{\cal F}_1^{(a)}(\lambda,\lambda_i).
%\nonumber \\
%&&
\ear
where 
the rapidity independent  constants
${\cal G}_{1}^{(a,b)}(\bar{\gamma})$ and
${\cal G}_{2}^{(a,b)}(\bar{\gamma})$ 
are given by
\bear
{\cal G}_{1}^{(a,b)}(\bar{\gamma})
&=&
\prod_{l=1}^{b-1}
\sqrt{
\frac{\sinh \left[\bar{\gamma} + \frac{\imath \pi k}{N} (a+l-2) \right] }
{\sinh \left[\bar{\gamma} + \frac{\imath \pi k}{N} (a+b-2) \right] }
\frac{\sinh \left[\frac{\imath \pi k}{N} (a+b-1-l) \right] }
{\sinh \left[\frac{\imath \pi k}{N} (a+b-1) \right] }
},
%\nonumber \\
%&&
\\
{\cal G}_{2}^{(a,b)}(\bar{\gamma})
&=&
\prod_{l=1}^{b-1}
\sqrt{
\frac{\sinh \left[\bar{\gamma} + \frac{\imath \pi k}{N} (a+l-1) \right] }
{\sinh \left[\bar{\gamma} + \frac{\imath \pi k}{N} (a-1) \right] }
\frac{\sinh \left[\frac{\imath \pi k}{N} (a+b-l) \right] }
{\sinh \left[\frac{\imath \pi k}{N} a \right] }
}.
\ear

The main ingredients to calculate the off-shell properties are completed by
presenting the explicit expressions for functions $R_{a,1}^{a,a}(\lambda,\mu)$
and $\theta(\lambda,\mu)$. For this model they are given by,
\EQ
R(\lambda,\mu)_{a,1}^{a,1}=
\prod_{j=1}^{a-1}
\frac{\sinh[\lambda-\mu+\frac{\imath \pi k}{N}(j-1) ]}{\sinh[\lambda-\mu+\bar{\gamma}+\frac{\imath \pi k}{N}(j-1)]},
\EN
and 
\EQ
\theta(\lambda,\mu)=
\frac{\sinh[\lambda-\mu-\frac{\imath \pi k}{N} ]}{\sinh[\lambda-\mu+\frac{\imath \pi k}{N}]}
\frac{\sinh[\lambda-\mu-\bar{\gamma} ]}{\sinh[\lambda-\mu+\bar{\gamma}]}.
\label{thetauni}
\EN

We would like to conclude this section with the following remarks.
Direct inspection of Eqs.(\ref{eigcol}-\ref{thetauni}) reveals us that for the
special point $\bar{\gamma}=-(N-1)\frac{\imath \pi k}{N}$ we are able to 
recover the corresponding
results (\ref{eigXXZ}-\ref{thetaXXZ}) concerning the solution of the 
XXZ-${\bf s}$ model with anisotropy $\gamma= -\frac{\pi k}{N}$.
We next note that the braid structure of the model at roots of unity
is  richer than that of the braid associated 
to the XXZ-${\bf s}$ model. In fact, by substituting
$\omega= q$ and $\bar{\gamma}=\imath(N-1)\gamma$ in Eq.(\ref{bracol1})  we
can reproduce the braid (\ref{braidXXZ}) 
of the XXZ-${\bf s}$ model up to a multiplicative normalization.  This means that at least formally
one can use such prescription in the Bethe ansatz results of this section to obtain the
corresponding ones
of the 
XXZ-${\bf s}$ model.

\subsection{Non-additive $R$-matrices}
\label{sec42}

Colored braid matrices can in general be thought as $R$-matrices depending on two independent
spectral parameters. In this case the braid carries two color variables $\lambda$ and $\mu$ 
attached to neighbor
strings which play the role of continuous variables. This two-parameter braid $\hat{S}(\lambda,\mu)$ satisfies the
following generalized braid relation,
\EQ
[I_N \otimes \hat{S}(\lambda_1,\lambda_2)][\hat{S}(\lambda_1,\lambda_3) \otimes I_N]
[I_N \otimes \hat{S}(\lambda_2,\lambda_3)]=
[\hat{S}(\lambda_2,\lambda_3) \otimes I_N][I_N \otimes \hat{S}(\lambda_1,\lambda_3)] 
[\hat{S}(\lambda_1,\lambda_2) \otimes I_N],
\label{bbcol}
\EN
where $\lambda_i$ are the color variables on the strings. 

It is immediate to see that solutions of the colored braid relation (\ref{bbcol})
provide us integrable vertex models with non-additive $R$-matrices 
$R(\lambda,\mu)$ upon the identification,
\EQ
R(\lambda,\mu)=PS(\lambda,\mu).
\EN

It turns out that the class of braid discussed in section (\ref{sec41}) 
admits such color extension \cite{DEG,DEG3,DEG1}. The color variables distinguish different
representations of $U_q[SU(2)]$ with dimension $N$ when $q$ is a root of unity \cite{COT1,SIE}. 
The simplest case $N=2$ is directly related to the Felderhof parameterization of the
free-fermion models \cite{FE}. Its $R$-matrix is that of a six-vertex model whose weights satisfy
the free-fermion condition, 
\bear
R(\lambda,\mu)&=&
(1-\lambda \mu)(e_{1,1}\otimes e_{1,1} + e_{2,2} \otimes e_{2,2})
+(\lambda-\mu) (e_{1,1} \otimes e_{2,2}-e_{2,2} \otimes e_{1,1})
\nonumber \\
&+&\sqrt{(1-\lambda^2)(1-\mu^2)}(e_{1,2}\otimes e_{2,1}+e_{2,1}\otimes e_{1,2}).
\ear

For $N \ge 3$ new vertex models start to emerge. 
The three-state case $N=3$ turns out to be an interesting nineteen-vertex model whose $R$-matrix
in our notation is,
\bear
R(\lambda,\mu)
&=&
R(\lambda,\mu)_{1,1}^{1,1}(e_{1,1}\otimes e_{1,1} + e_{3,3} \otimes e_{3,3})
+
R(\lambda,\mu)_{1,2}^{1,2}(e_{1,1}\otimes e_{2,2} - e_{2,2} \otimes e_{1,1})
\nonumber \\
&+&
R(\lambda,\mu)_{1,2}^{2,1}(e_{1,2}\otimes e_{2,1} + e_{2,1} \otimes e_{1,2})
+
R(\lambda,\mu)_{1,3}^{1,3} e_{1,1}\otimes e_{3,3}
\nonumber \\
&+&
R(\lambda,\mu)_{1,3}^{2,2}(e_{1,2}\otimes e_{3,2} + e_{2,1} \otimes e_{2,3})
+
R(\lambda,\mu)_{1,3}^{3,1} ( e_{1,3}\otimes e_{3,1} + e_{3,1}\otimes e_{1,3})
\nonumber \\
&+&
R(\lambda,\mu)_{2,2}^{2,2} e_{2,2}\otimes e_{2,2}
+
R(\lambda,\mu)_{2,2}^{3,1}(e_{2,3} \otimes e_{2,1} + e_{3,2}\otimes e_{1,2})
\nonumber \\
&+&
R(\lambda,\mu)_{2,3}^{2,3}(e_{2,2}\otimes e_{3,3} - e_{3,3} \otimes e_{2,2})
+
R(\lambda,\mu)_{2,3}^{3,2}(e_{2,3} \otimes e_{3,2} + e_{3,2}\otimes e_{2,3})
\nonumber \\
&+&
R(\lambda,\mu)_{3,1}^{3,1} e_{3,3} \otimes e_{1,1},
\ear
where the Boltzmann weights amplitudes $R(\lambda,\mu)_{a,b}^{c,d}$ are given by \cite{DEG}
\bear
R(\lambda,\mu)_{1,1}^{1,1}&=&(1-\mu \lambda)(1-\mu \lambda w),
\\
R(\lambda,\mu)_{1,2}^{1,2}&=&(\lambda-\mu)(1-\mu \lambda w),
\\
R(\lambda,\mu)_{1,2}^{2,1}&=&(1-\mu \lambda w) \sqrt{(1-\mu^2)(1-\lambda^2)},
\\
R(\lambda,\mu)_{1,3}^{1,3}&=&(\lambda-\mu)(\lambda-\mu w),
\\
R(\lambda,\mu)_{1,3}^{2,2}&=&(\lambda-\mu)\sqrt{(1-\lambda^2)(1-\mu^2 w)(1+w)},
\\
R(\lambda,\mu)_{1,3}^{3,1}&=& \sqrt{(1-\lambda^2)(1-\lambda^2 w)(1-\mu^2)(1-\mu^2 w)},
\\
R(\lambda,\mu)_{2,2}^{2,2}&=&(1-\lambda^2)(1-\mu^2 w)-(\mu-\lambda)(\mu-\lambda w),
\ear
\bear
R(\lambda,\mu)_{2,2}^{3,1}&=&(\mu-\lambda) \sqrt{(1-\mu^2)(1-\lambda^2 w)(1+w)},
\\
R(\lambda,\mu)_{2,3}^{2,3}&=&(1+w)(\lambda-\mu)(1-\mu \lambda),
\\
R(\lambda,\mu)_{2,3}^{3,2}&=&(1-\mu \lambda) \sqrt{(1-\mu^2 w)(1-\lambda^2 w)},
\\
R(\lambda,\mu)_{3,1}^{3,1}&=&(\mu-\lambda)(\mu-\lambda w),
\ear
such that $\omega=\exp(\frac{2 \pi \imath}{3})$ or
$\omega=-\exp(\frac{ \pi \imath}{3})$. 

For this family of vertex models the expression for the respective $R$-matrix are rather 
involving for arbitrary $N$ \cite{DEG,DEG3,DEG1,SIE}. In the case $N=4$,  we recall that 
the explicit formulae 
for all the weights  $R_{a,b}^{c,d}(\lambda,\mu)$  are available in \cite{DEG} 
and for sake of completeness we include them
in Appendix B. Fortunately, by substituting the $R$-matrix amplitudes 
for $N=2,3,4$ in Eqs.(\ref{beaN},\ref{gaman}), 
we noticed that such 
on-shell Bethe ansatz results have a uniform dependence on $N$. This makes possible to propose 
the transfer matrix  eigenvalues for general $N$, 
\EQ
\Lambda_{n}(\lambda) =
\sum_{a=1}^{N}
\prod_{j=1}^{L}
\left[
\prod_{i=a}^{N-1}(1-\lambda \mu_j w^{i-1})
\prod_{i=1}^{a-1}(\mu_j-\lambda w^{i-1})
\right]
\prod_{j=1}^{n}
\frac{(1-\lambda \lambda_j)(\lambda-\lambda_j w)}{\lambda w^{a-1}-\lambda_j}\frac{w^{a-2}}{\lambda w^{a-2}-\lambda_j},
\EN
while the Bethe ansatz equations for the variables $\lambda_j$ are,
\EQ
\prod_{j=1}^{L}
\frac{1-\lambda_l \mu_j}{\mu_j-\lambda_l} =
\prod_{\stackrel{j=1}{j \neq l}}^{n}
\frac{\lambda_l-\lambda_j w}{\lambda_l w-\lambda_j},
~~l = 1, \dots, n.
\EN

The same observation made above also works for the off-shell properties.
A case by case analysis of Eqs.(\ref{psi1}-\ref{fbb}) up to $N=4$, following
the strategy of Appendix A,
is able to reveal us the general pattern for functions
${_0}{\cal F}_b^{(a)}(\lambda,\lambda_1,\dots,\lambda_b)$
and ${_b}{\cal F}_b^{(a)}(\lambda,\lambda_1,\dots,\lambda_b)$, namely 
\EQ
{_0}{\cal F}_1^{(a)}(\lambda,\mu)
=
-{_1}{\cal F}_1^{(a)}(\lambda,\mu)=
\frac{\sqrt{(1-w^a) (1-w^{a-1} \lambda^2)(1-\mu^2)}}{\sqrt{1-w} (\mu-w^{a-1} \lambda)},
\EN
\bear
{_0}{\cal F}_b^{(a)}(\lambda,\lambda_1,\dots,\lambda_b)
&=&
w^{\frac{b(b-1)}{2}}
\prod_{\stackrel{i,j=1}{i<j}}^{b}
\frac{(1-\lambda_i \lambda_j)}{(w \lambda_i-\lambda_j)}
\prod_{i=1}^{b-1}
\sqrt{
\left( \frac{1-w^{a+b-1-i}}{1-w^{a+b-1}}\right)
\left(\frac{1-\lambda^2 w^{a+i-2}}{1-\lambda^2 w^{a+b-2}} \right)}
\nonumber \\
&\times&
\prod_{i=1}^{b}
{_0}{\cal F}_1^{(a+b-1)}(\lambda,\lambda_i),
\ear
\bear
{_b}{\cal F}_b^{(a)}(\lambda,\lambda_1,\dots,\lambda_b)
&=&
\prod_{\stackrel{i,j=1}{i<j}}^{b}
\frac{(1-\lambda_i \lambda_j)}{(w \lambda_i-\lambda_j)}
\prod_{i=1}^{b-1}
\sqrt{
\left(\frac{1-w^{a+b-i}}{1-w^{a}} \right)
\left(\frac{1-\lambda^2 w^{a+i-1}}{1-\lambda^2 w^{a-1}} \right)}
\nonumber \\
&\times&
\prod_{i=1}^{b}
{_1}{\cal F}_1^{(a)}(\lambda,\lambda_i).
\ear

As before the off-shell amplitudes are completely determined by presenting  
the respective functions $R_{a,1}^{a,1}(\lambda,\mu)$
and $\theta(\lambda,\mu)$. For this model they are,
\EQ
R_{a,1}^{a,1}(\lambda,\mu)
=
\prod_{i=1}^{a-1} \left(\mu-\lambda w^{i-1} \right)
\prod_{i=a}^{N-1} \left(1-\mu \lambda w^{i-1} \right),
\EN
and
\EQ
\theta(\lambda,\mu)
=
- \frac{\lambda-\mu w}{\lambda w-\mu}.
\EN

We close this section by mentioning that a colored vertex model with an infinite number
of edge states has also been proposed by
Deguchi and Akutsu \cite{DEG3,DEG1}. This system is
invariant by the $U(1)$ symmetry and therefore can in principle be solved
within the algebraic Bethe ansatz approach developed by the authors of this
paper \cite{CM}. We hope to address the problem  of presenting the on-shell and
off-shell Bethe ansatz properties of this model in a future publication. 

\section{Non-Compact $SL(2,{\cal R})$ model}

In this section we present the on-shell and off-shell algebraic Bethe ansatz properties of 
the vertex model based on the discrete $D^{-}_{\bf{s}}$ representation of the $SL(2,{\cal R})$ algebra.
The interest in the study of such non-compact vertex models emerged from the 
discovery of integrable structures in high energy QCD scattering \cite{LI,FAKO,DEV}.
It has been argued that the scale dependence of certain scattering amplitudes 
is governed by the eigenspectrum of exactly solvable Hamiltonians with $SL(2,{\cal R})$ symmetry \cite{BDM}.
In recent years similar connection was found in the context of the duality 
between ${\cal N}=4$ Yang-Mills gauge theory and the world-sheet theory of strings 
on $AdS^5 \times S^5$ \cite{MZ,MZ1}.
In particular, the simplest non-compact subsector of the one-loop 
dilatation operator of the ${\cal N}=4$ gauge theory is directly 
related to the ${\bf{s}}=-\frac{1}{2}$ integrable $SL(2,{\cal R})$ spin magnet \cite{BEI}.

We start by recalling that the $SL(2,{\cal R})$ algebra is generated by 
three operators obeying the following commutation rules,
\EQ
[S^z,S^{\pm}]=\pm S^{\pm}, ~~~[S^+,S^-]=2 S^z.
\label{ALG}
\EN

Here we will consider integrable models associated to the discrete highest weight $D^{-}_{{\bf{s}}}$
representation of $SL(2,{\cal R})$.
This representation is labeled by the generalized spin variables 
such that ${\bf{s}} \in {\cal R}^{-}$ for the universal covering group $SL(2,{\cal R})$ \cite{BARU}.
The respective states can be represented in terms of the following infinite dimensional angular momenta basis,
\EQ
\ket{{\bf{s}},m+{\bf{s}}}, ~~m=0,1,\dots,\infty,
\label{STA}
\EN
where $S^z \ket{{\bf{s}},m+{\bf{s}}}=(m+{\bf{s}}) \ket{{\bf{s}},m+{\bf{s}}}$ 
and $S^+ \ket{{\bf{s}},{\bf{s}}}=0$.
The state $\ket{{\bf s},{\bf s}}$ plays therefore the role of a highest weight vector.

The simplest $R$-matrix associated to the $SL(2,{\cal R})$ algebra has a two dimensional auxiliary space.
It can be viewed as a $2 \times 2$ matrix given by \cite{RKS,FAT}
\EQ
R_{\frac{1}{2},{\bf{s}}}(\lambda,\mu)=\left(\begin{array}{cc}
                (\lambda-\mu)I_{\bf{s}} + \imath S^z & \imath S^- \\
                \imath S^+ & (\lambda-\mu)I_{\bf{s}}-\imath S^z \\
                \end{array}\right),
\label{RMA}
\EN
where $I_{\bf{s}}$ denotes an infinite dimensional unity matrix.

However, to construct non-compact spin magnets described by next-neighbor 
Hamiltonians we have to consider a $R$-matrix whose auxiliary space belongs to the infinite dimensional representation $D^{-}_{{\bf{s}}}$.
Such $R$-matrix turns out to be a non-trivial generalization of (\ref{RMA}) and has the following form \cite{RKS},
\EQ
R_{{\bf{s}},{\bf{s}}}(\lambda,\mu)
=
\sum_{j=0}^{\infty} \prod_{k=1}^j
\left(\frac{k \imath+\lambda-\mu}{k\imath - \lambda + \mu} \right) \check{P}_j({\bf{s}}),
\label{RMAS}
\EN
where $\check{P}_j({\bf{s}})$ are operators projecting 
the tensor product space $D^{-}_{\bf{s}} \otimes D^{-}_{\bf{s}}$ on the representation with total spin $j$.

The corresponding Hamiltonian can be derived by taking the logarithmic 
derivative of the transfer matrix (\ref{tran},\ref{mono1},\ref{lope}) whose 
respective weights $R(\lambda,\mu)_{a,b}^{c,d}$ are
obtained from the $R$-matrix (\ref{RMAS}).
We recall that this derivative is computed at the regular point $\lambda=0$ and also by 
setting the inhomogeneities $\mu_l$ to zero.
The Hamiltonian is then given by the standard expression,
\EQ
H
=
\sum_{l=1}^{L-1} H_{i,i+1}({\bf{s}})+H_{L,1}({\bf{s}}),
\EN
where $H_{1,2}({\bf s})= \frac{d}{d \lambda} ln R_{{\bf s},{\bf s}}(\lambda,0)|_{\lambda=0}$.

As far as we know the explicit expression for density 
Hamiltonian $H_{1,2}({\bf s})$ has only been discussed in the particular case ${\bf s}=-\frac{1}{2}$ \cite{BEI}.
In what follows we shall complement this result by presenting the 
action of $H_{1,2}({\bf s})$ on the tensor product of states (\ref{STA}) for arbitrary ${\bf s} \in {\cal R}^{-}$.
We emphasize that this knowledge is essential for practical applications of such non-compact spin chains.
The computation is somehow involving but the final result is rather simple \footnote{We have set the
overall normalization such that  
$H_{1,2}({\bf s}) \ket{0,0}=0$.},
\bear
H_{1,2}({\bf s}) \ket{m_1,m_2}
&=&
\left[
\sum_{k=1}^{m_1} h_1(k)
+
\sum_{k=1}^{m_2} h_1(k)
\right]
\ket{m_1,m_2}
\nonumber \\
&+&
\sum_{k=1}^{m_1} h_2(k,m_2,m_1) \ket{m_1-k,m_2+k}
\nonumber \\
&+&
\sum_{k=1}^{m_2} h_2(k,m_1,m_2) \ket{m_1+k,m_2-k},
\label{hnon}
\ear
where $\ket{m_1,m_2}$ denotes a sort notation for the tensor product $\ket{{\bf s},{\bf s}+m_1} \otimes \ket{{\bf s},{\bf s}+m_2}$ state.
In addition, functions $h_1(k)$ and $h_2(k,m_1,m_2)$ 
are given by \footnote{Note that for ${\bf s}=-\frac{1}{2}$ 
functions $h_1(k)$ and $h_2(k,m_1,m_2)$ drastically simplify to the form $\pm \frac{1}{k}$ of harmonic numbers.}
\bear
h_1(k) &=& \frac{2 {\bf s}}{2{\bf s}+1-k},
\\
h_2(k,m_1,m_2) &=& \frac{2 {\bf s}}{k} \prod_{i=1}^{k} \sqrt{\frac{(m_1+i)}{(m_1+i-2{\bf s}-1)} \frac{(m_2+1-i)}{(m_2-2{\bf s}-i)}}.
\label{hnon1}
\ear

At the present the algebraic Bethe ansatz analysis of the vertex models has 
been restricted to the diagonalization of the transfer matrix whose weights are 
the elements of the simplest $R$-matrix $R_{\frac{1}{2},{\bf s}}(\lambda,\mu)$ \cite{FAKO,SL2}.
In this case the auxiliary space is two-dimensional and the algebraic 
Bethe ansatz is fairly  parallel to that 
developed for six-vertex models \cite{FA,KO}.
The same problem for the transfer matrix based on the 
$R$-matrix $R_{{\bf s},{\bf s}}(\lambda,\mu)$ could in principle be 
handled by extending 
the approach used to solve the $SU(2)$ higher 
spin XXX-${\bf s}$ chain \cite{BAB1} to the 
situation of an infinite-dimensional auxiliary space.
This method, however, is not capable to keep track of all the unwanted 
terms and to find the Bethe equations from the condition of their equality to zero.
This drawback prevents us to find a complete algebraic Bethe ansatz 
solution and consequently to obtain information on the off-shell properties.

By way of contrast we certainly can use the results of section 2 to 
tackle the  problem for the transfer matrix constructed from the $R$-matrix $R_{{\bf s},{\bf s}}(\lambda,\mu)$.
The suitable reference state for this model is constructed by considering 
the tensor product of highest vector $\ket{{\bf s},{\bf s}}$,
\EQ
\ket{0}= \prod_{i=1}^{L}  
\otimes \ket{{\bf s},{\bf s}}_i.
\EN

Our next task is to compute the $R$-matrix elements 
of $R_{{\bf s},{\bf s}}(\lambda,\mu)$ that are necessary to 
calculate the corresponding on-shell and off-shell properties.
In general this task is rather complicated since both 
the auxiliary and quantum spaces of (\ref{RMAS}) are infinite dimensional.
This problem can be circumvented by exploring the $U(1)$ invariance and 
expressing the $R$-matrix in terms of the sectors $n$ 
labeling the eigenvalues of the $U(1)$ operator. More precisely,
we can always decompose 
$R_{{\bf s},{\bf s}}(\lambda,\mu)$ as, 
\EQ
R_{{\bf s},{\bf s}}(\lambda,\mu) =\sum_{n=0}^{\infty}
\sum_{a,c=1}^{n + 1}
R(\lambda,\mu)_{a,n+2-a}^{c,n+2-c}
e_{a,c}
\otimes
e_{n+2-a,n+2-c}.
\label{RMAS1}
\EN

We now  project out the operators
${\check P}_j(s)$ on a given sector $n$ and by comparing 
Eqs.(\ref{RMAS},{\ref{RMAS1}) we are able to calculate
the weights
$R(\lambda,\mu)_{a,n+2-a}^{c,n+2-c}$.   
As concrete examples of this approach we have summarized in Appendix C such amplitudes up 
to sector $n=4$ for arbitrary value of ${\bf s} \in {\cal R}$.
By substituting such elements in Eq.(\ref{beaN},\ref{gaman}) and 
carrying on some algebraic simplifications 
one observes that each transfer matrix eigenvalue term has a very simple dependence on sector $n$.
This enables us to propose the exact expression for the eigenvalues,
\bear
\Lambda_n(\lambda) &=&
\sum_{a=1}^{\infty}
\prod_{l=1}^{L}
\prod_{k=1}^{a-1}
\frac{[\lambda-\mu_l-(k-1)\imath ]}{[\lambda-\mu_l+(2{\bf s}+1-k)\imath ]}
\nonumber \\
& \times & \prod_{i=1}^{n}
\frac{[\lambda-\lambda_i-2 {\bf s} \imath] [\lambda-\lambda_i+\imath ]}
{[\lambda-\lambda_i- (a-1)\imath] [\lambda-\lambda_i- (a-2)\imath]}.
\label{eigXXX}
\ear

The corresponding Bethe ansatz equations are fixed from the knowledge of the $R$-matrix elements
up to the sector $n=2$.
We find that the rapidities $\lambda_j$ satisfy the following equation,
\footnote{The Bethe ansatz
equations (\ref{beaXXX}) can be  symmetrized through the
shift $\lambda_j \rightarrow
\lambda_j -{\bf s}\imath $.}, 
\EQ
\prod_{l=1}^{L}
\frac{(\lambda_j-\mu_l+ 2{\bf s}\imath )}{(\lambda_j-\mu_l)}
 =
\prod_{\stackrel{i=1}{i \neq j}}^{n}
\frac{(\lambda_j-\lambda_i+\imath )}{(\lambda_j-\lambda_i-\imath)}
,~~~j = 1, \dots, n. 
\label{beaXXX}
\EN

Before proceeding we remark  
that Eq.(\ref{beaXXX}) can be reproduced within
the coordinate Bethe ansatz formulation for the non-compact 
Hamiltonian (\ref{hnon}-\ref{hnon1}).  This has been done 
in the very special case of the spin ${\bf s}=-\frac{1}{2}$ up to
the two particle excitation sector \cite{STAU}. For sake of completeness we have
presented an extension of this analysis for arbitrary values of ${\bf s}$ in Appendix D.
In the case of integrable
theories this is enough to supply us the main form of
the Bethe equations for the rapidities but not 
sufficient to provide us the general structure
of the eigenvectors. The situation is even more complicated for
non-compact systems due to the possibility of an infinite number of
particle excitations per site. 

However, an algebraic representation for the 
eigenvectors of the non-compact vertex model can formally be  
obtained by taking the limit
$ N \rightarrow \infty $ in Eq.(\ref{eing}).  Therefore, 
to benefit from the knowledge of the eigenvectors we have to
compute the off-shell functions
${_b}{\cal F}_b^{(2)}(\lambda,\lambda_1,\cdots,\lambda_b)$ where now
the indices $a$ and $b$ are unlimited. Fortunately, 
this computation
can be implemented by using the same strategy explained above for the 
on-shell data. It turns out that the final results for the off-shell
properties are,  
\EQ
{_0}{\cal F}_1^{(a)}(\lambda,\mu)
=
-{_1}{\cal F}_1^{(a)}(\lambda,\mu)
=-\imath
\frac{\sqrt{ 2 {\bf s} a (2 {\bf s} +1-a)}}{\mu-\lambda +(a-1)\imath}
\label{f1xxx}
\EN
\bear
{_0}{\cal F}_b^{(a)}(\lambda,\lambda_1,\dots,\lambda_b)
&=&
\prod_{k=1}^{b-1} \sqrt{\frac{(2 {\bf s} +2 -a -k)(a+b-1-k)}{
(2 {\bf s} +2 -a -b)(a+b-1)}}
\prod_{\stackrel{i,j=1}{i<j}}^{b}
\frac{( \lambda_i-\lambda_j -2 {\bf s}\imath )}{( \lambda_i-\lambda_j-\imath )}
\nonumber \\
& \times & \prod_{i=1}^{b}
{_0}{\cal F}_1^{(a+b-1)}(\lambda,\lambda_i) ,
%\nonumber
\\
{_b}{\cal F}_b^{(a)}(\lambda,\lambda_1,\cdots,\lambda_b)
&=&
\prod_{k=1}^{b-1} \sqrt{\frac{(2 {\bf s} +1 -a -k)(a+b-k)}{
(2 {\bf s} +1 -a)a}}
\prod_{\stackrel{i,j=1}{i<j}}^{b}
\frac{( \lambda_i-\lambda_j -2{\bf s} \imath)}{( \lambda_i-\lambda_j-\imath )}
\nonumber \\
& \times &
\prod_{i=1}^{b}
{_1}{\cal F}_1^{(a)}(\lambda,\lambda_i),
\label{f3xxx}
\ear
where $a,b=1,\dots,\infty$.

The off-shell data is completed by exhibiting functions 
$R_{a,1}^{a,1}(\lambda,\mu)$ and $\theta(\lambda,\mu)$. 
For such non-compact vertex model they are,
\EQ
R(\lambda,\mu)_{a,1}^{a,1}=\prod_{k=1}^{a-1}
\frac{[\lambda-\mu- (k-1)\imath]}{[\lambda-\mu+ (2 {\bf s }+1-k)\imath]},
\EN
and 
\EQ
\theta(\lambda,\mu)=
\frac{(\lambda-\mu-2 \bf{s}\imath )}{(\lambda-\mu+2 \bf{s} \imath)}
\frac{(\lambda-\mu+\imath )}{(\lambda-\mu-\imath)}.
\label{f5xxx}
\EN

At this point we note that the Bethe ansatz properties of the non-compact
$SL(2,{\cal R})$ model  
can be viewed as an analytic continuation of those
associated to the XXX-${\bf s}$ model. We just have to extend the
spin variable to take values on ${\bf s} \in 
{\cal R}^{-}$ as well as to consider the total number of degrees
of freedom per site infinite.  This feature can be seen by considering
the isotropic limit $\gamma \rightarrow 0$ in the results for the XXZ-${\bf s}$
and afterwards comparing them with the structure of 
Eqs.(\ref{eigXXX}-\ref{f5xxx}).  This property, to what concerns the form of the Bethe
ansatz equations, was expected from the algebraic Bethe ansatz for the
$R$-matrix $R_{\frac{1}{2},{\bf s}}(\lambda,\mu)$ \cite{FAKO,SL2}. The fact that it extends
to the off-shell amplitudes is however a novelty which strengths the relationship between 
representation theory and Bethe ansatz properties.

Other remarkable feature of the on-shell and off-shell results for the
non-compact model is as follows. It turns out that they can also be
obtained from a particular limit of those 
derived in section \ref{sec41} for the vertex model based on the quantum algebra
at roots of unity.  This is achieved by choosing the free parameter 
$\bar{\gamma}= -2 {\bf s} \frac{\imath \pi k}{N}$ and also by re-scaling all the
spectral variables as follows 
$\lambda \rightarrow -\frac{\pi k}{N} \lambda$,  
$\lambda_j \rightarrow -\frac{\pi k}{N} \lambda_j$ and  
$\mu_l \rightarrow -\frac{\pi k}{N} \mu_l$.  
By performing these operations in Eqs.(\ref{eigcol}-\ref{thetauni})
and afterwards taking the 
$N \rightarrow \infty $ limit we indeed 
obtain the results (\ref{eigXXX}-\ref{f5xxx})
associated to the non-compact model.  This fact offers
us the possibility to study the properties of such non-compact model considering
a well defined limit of a system with finite number of degrees of freedom. This
truncated approach can be seen as one way to infer on the 
physical behavior of the non-compact model avoiding the complications of dealing with
infinite dimensional Hilbert space. We conclude by mentioning that such mechanism also works
if we use as a compact system
the XXZ-${\bf s}$ model. In this case we have to tune the spin and anisotropy by choosing
for example
$\gamma= \frac{2 \pi}{N}$. We then take the
limit $N \rightarrow \infty$ and as a result we are able to recover the on-shell and off-shell behavior
associated to the non-compact ${\bf s} =-\frac{1}{2}$ chain. From the higher 
spin Heisenberg model, however,
the possible values we can reach for the non-compact spin variable ${\bf s}$ are restricted.

\section{Conclusions}

We have presented the algebraic Bethe ansatz solution of three distinct classes of
integrable vertex models that are invariant by one $U(1)$ charge symmetry. The
on-shell and off-shell properties associated to the respective transfer matrix
eigenvalue problems are exhibited. In particular, all the off-shell data
can be presented in term of products of elementary building block functions. 
This fact could be of relevance to compute properties
that require the knowledge of the exact form of the eigenvectors such
wave-function norms and correlation functions of $U(1)$ higher spin
chains \cite{KUT,DEGC}.

The first two families of vertex models are derived from the braid
group representations of the quantum $U_q[SU(2)]$ algebra 
either for generic values
of the deformation parameter producing the XXZ-${\bf s}$ chain or when it
takes values on the roots of unity leading us to colored models. It is noted
that the latter braid  representation is richer than the one associated to
generic values of $q$. Formally, this property allows one to obtain the Bethe
ansatz results for the XXZ-${\bf s}$ chain by adapting those derived
for the colored vertex model. Here we remark that recently these solvable
models have been discussed in the context of the calculation
of their partition function on the presence of certain domain wall boundary
condition \cite{CARA}. Considering that this kind of partition functions
can in principle be calculated by the algebraic Bethe ansatz method \cite{PRO}
it seems interesting to investigate whether the results of \cite{CARA}
can be reproduced or even extended to other possible fix
boundary conditions by using the algebraic Bethe ansatz framework described
in this paper. 

The third class of model considered is that based on the discrete
$D_{{\bf s}}^{-}$ representation of the $SL(2,{\cal R})$ algebra.
We have derived the expression for the corresponding Hamiltonian 
acting on the standard angular momenta basis. This provides us
the means to derive the coherent state representation of this
integrable spin chain and to find in the continuum limit the
respective two-dimensional quantum field theory. Both on-shell
and off-shell Bethe ansatz results for such non-compact model can be seen
as a kind of analytic continuation of those associated to the
XXZ-${\bf s}$. In addition, we argued that these Bethe ansatz
properties can be derived from the diagonalization of the transfer
matrix of the $U_q[SU(2)]$ vertex model at roots of unity. This
observation offers us the possibility to unveil the physical
behavior of the non-compact $SL(2, {\cal R})$ model from
a bona fide $N \rightarrow \infty$ limit of the properties
of a compact vertex model. This avoids us to deal with
infinite dimensional Hilbert space specially to what
concerns the study of the  anti-ferromagnetic behavior
of the $SL(2,{\cal R})$ chain.  

It is reasonable to believe that the above remarks
are not restricted to the $q$-deformation of the classical
$SU(2)$ symmetry. In fact, the existence of colored vertex
models associated to quantum groups other than
$U_q[SU(2)]$ has already been outlined 
in the literature \cite{DEG3}.
This then could supply us with a general method
to extract information about an integrable non-compact
model based on a given algebra  ${\cal G}$ from the respective
compact model associated to the $U_q[{\cal G}]$ deformation
at roots of unity. It would be rather interesting to
investigate this fact in the case of superalgebras since
an extra continuum variable besides the deformation parameter
is allowed \cite{GUS}. In particular, if this approach could bring any new
insight to exactly solved models based on 
the non-compact $PSU(2,2|4)$ algebra due
to their apparent relevance in the understanding
of the integrable properties found for planar ${\cal N} =4$ super
Yang-Mills theory \cite{MZ,MZ1,BEI}.

\section*{Acknowledgments}
The authors thank the Brazilian Research Agencies FAPESP and CNPq for financial support.

\section*{\bf Appendix A : Off-shell Amplitudes }
\setcounter{equation}{0}
\renewcommand{\theequation}{A.\arabic{equation}}

In this appendix we present the technical details concerning the computation
of the off-shell amplitudes
${_0}{\cal F}_b^{(a)}(\lambda,\lambda_1,\cdots,\lambda_b)$
and ${_b}{\cal F}_b^{(a)}(\lambda,\lambda_1,\cdots,\lambda_b)$. We first note
that for $b=1$ such 
amplitudes are directly computed from the knowlodge of the Boltzmann weigths
by using Eq.(\ref{psi1}). For $b \ge 2$ we have to iterate the
recurrence relations (\ref{fbc}-\ref{fbb}) for all the possible values
of the index $c=0,\cdots,b$. In the simplest case $b=2$ such equations
lead us to the expressions,
\bear
\label{f21}
{_1}{\cal F}_{2}^{(a)}(\lambda,\lambda_1,\lambda_2)&=&
{_0}{\cal F}_{1}^{(a)}(\lambda,\lambda_2)
{_1}{\cal F}_{1}^{(a+1)}(\lambda,\lambda_1)
\frac{R(\lambda_2,\lambda_1)_{1,1}^{1,1}}{R(\lambda_2,\lambda_1)_{2,1}^{2,1}}
\\
\label{f20}
{_0}{\cal F}{_2^{(a)}}(\lambda,\lambda_1,\lambda_2)&=&
\frac{R(\lambda,\lambda_1)_{a+1,1}^{a,2}}{R(\lambda,\lambda_1)_{a+2,1}^{a+2,1}}
{_0}{\cal F}_{1}^{(a+1)}(\lambda,\lambda_2)+
\frac{R(\lambda,\lambda_1)_{a+2,1}^{a,3}}{R(\lambda,\lambda_1)_{a+2,1}^{a+2,1}}
{_1}{\cal F}_{1}^{(2)}(\lambda_1,\lambda_2)
\\
\label{f22}
{_2}{\cal F}{_2^{(a)}}(\lambda,\lambda_1,\lambda_2)
&= &
-{_0}{\cal F}{_2^{(a)}}(\lambda,\lambda_1,\lambda_2)
-
{_1}{\cal F}{_2^{(a)}}(\lambda,\lambda_2,\lambda_1) 
\frac{R(\lambda_1,\lambda_2)_{2,1}^{2,1}}{R(\lambda_1,\lambda_2)_{1,1}^{1,1}} 
\frac{R(\lambda_2,\lambda_1)_{1,1}^{1,1}}{R(\lambda_2,\lambda_1)_{2,1}^{2,1}}
\nonumber \\
&-&{_1}{\cal F}{_2^{(a)}}(\lambda,\lambda_1,\lambda_2) 
\frac{R(\lambda_2,\lambda_1)_{2,1}^{2,1}}{R(\lambda_2,\lambda_1)_{1,1}^{1,1}} 
\frac{R(\lambda_1,\lambda_2)_{1,1}^{1,1}}{R(\lambda_1,\lambda_2)_{2,1}^{2,1}}
\theta(\lambda_1, \lambda_2). 
\ear

From Eq.(\ref{f21}) we see that 
${_1}{\cal F}_{2}^{(a)}(\lambda,\lambda_1,\lambda_2)$ is already given in terms
of product of the one-particle $b=1$ off-shell amplitudes and the ratio
of elementary weights. Therefore, we only have to carry on simplifications
on the amplitudes 
${_0}{\cal F}{_2^{(a)}}(\lambda,\lambda_1,\lambda_2)$ and
${_2}{\cal F}{_2^{(a)}}(\lambda,\lambda_1,\lambda_2)$ associated to the
extremum values of the index $c=0,2$.  This step is done by substituting in
Eqs.(\ref{f20},\ref{f22}) the previous results for the $b=1$ off shell amplitudes
as well as the expression for ${_1}{\cal F}{_2^{(a)}}(\lambda,\lambda_1,\lambda_2)$.  
It turns out that
for all the models considered in this paper we find that such amplitudes 
can be presented in the following factorized form,
\EQ
{_0}{\cal F}_b^{(a)}(\lambda,\lambda_1,\lambda_2)
= {\cal{A}}_0^{(a,2)}
Q(\lambda_1,\lambda_2) 
\prod_{i=1}^{2} 
{_0}{\cal F}{_1^{(a+1)}}(\lambda,\lambda_i)
\label{ans1}
\EN
and
\EQ
{_2}{\cal F}_b^{(a)}(\lambda,\lambda_1,\lambda_2)
= {\cal{A}}_1^{(a,2)}
Q(\lambda_1,\lambda_2) 
\prod_{i=1}^{2} 
{_1}{\cal F}{_1^{(a)}}(\lambda,\lambda_i)
\label{ans2}
\EN

The parameters ${\cal{A}}_0^{(a,2)}$ 
and ${\cal{A}}_1^{(a,2)}$  are constants which depend of the model we are
considering but they are 
independent of the variables $\lambda$,$\lambda_1$ and $\lambda_2$. In addition, 
function
$Q(\lambda,\mu)$ depends on the corresponding weights by an expression
that is model independent, namely
\EQ
Q(\lambda,\mu)= \theta(\lambda,\mu) 
\frac{R(\lambda,\mu)_{1,1}^{1,1}}{R(\lambda,\mu)_{2,1}^{2,1}} 
+\frac{R(\mu,\lambda)_{1,1}^{1,1}}{R(\mu,\lambda)_{2,1}^{2,1}} 
\EN
where $\theta(\lambda,\mu)$ is determined 
Eq.(\ref{theta}). 

The factorized form of all off-shell amplitudes associated to
the two-particle sector can now be used to determine the structure
of the off-shell amplitudes for the next sector $b=3$.  This is once again
done
by iterating the recurrence relations (\ref{fbc}-\ref{fbb}). By performing
this procedure up to the four-particle sector we conclude that 
the structure of functions  
${_0}{\cal F}_b^{(a)}(\lambda,\lambda_1,\cdots,\lambda_b)$
and ${_b}{\cal F}_b^{(a)}(\lambda,\lambda_1,\cdots,\lambda_b)$ are
given by the expressions,
\EQ
{_0}{\cal F}_b^{(a)}(\lambda,\lambda_1,\dots,\lambda_b)
= {\cal A}_0^{(a,b)} 
\prod_{\stackrel{i,j=1}{i<j}}^{b} Q(\lambda_i,\lambda_j)
\prod_{i=1}^{b}
{_0}{\cal F}_1^{(a+b-1)}(\lambda,\lambda_i) ,
\label{anti1}
\EN
and 
\EQ
{_b}{\cal F}_b^{(a)}(\lambda,\lambda_1,\cdots,\lambda_b)
=
{\cal A}_1^{(a,b)}
\prod_{\stackrel{i,j=1}{i<j}}^{b} Q(\lambda_i,\lambda_j)
\prod_{i=1}^{b}
{_1}{\cal F}_1^{(a)}(\lambda,\lambda_i),
\label{anti2}
\EN
where
${\cal A}_0^{(a,b)}$ and  
${\cal A}_1^{(a,b)}$ are model dependent constants. 

Having found the functional 
dependence of the off-shell functions on the
rapidities the next task is to determine the constants 
${\cal A}_0^{(a,b)}$ and  
${\cal A}_1^{(a,b)}$.  This is done by direct comparison of
explicit calculations for functions 
${_0}{\cal F}_b^{(a)}(\lambda,\lambda_1,\cdots,\lambda_b)$
and ${_b}{\cal F}_b^{(a)}(\lambda,\lambda_1,\cdots,\lambda_b)$ 
with the expression given by Eqs.(\ref{anti1},\ref{anti2}).
The explicit computation 
of such constants are more cumbersome for the 
colored vertex model due to the many possible branches.

\section*{\bf Appendix B : Non-additive $N=4$ $R$-matrix}
\setcounter{equation}{0}
\renewcommand{\theequation}{B.\arabic{equation}}

Here we rewrite the colored $R$-matrix for $N=4$ given in \cite{DEG} considering
the notation used in this paper. The result is,
\bear
R_{12}(\lambda,\mu)
&=&
R(\lambda,\mu)_{1,1}^{1,1}(e_{1,1}\otimes e_{1,1} + e_{4,4} \otimes e_{4,4})
+
R(\lambda,\mu)_{1,2}^{1,2}(e_{1,1}\otimes e_{2,2} - e_{2,2} \otimes e_{1,1})
\nonumber \\
&+&
R(\lambda,\mu)_{1,2}^{2,1}(e_{1,2}\otimes e_{2,1} + e_{2,1} \otimes e_{1,2})
+
R(\lambda,\mu)_{1,3}^{1,3} e_{1,1}\otimes e_{3,3}
\nonumber \\
&+&
R(\lambda,\mu)_{1,3}^{2,2}(e_{1,2}\otimes e_{3,2} + e_{2,1} \otimes e_{2,3})
+
R(\lambda,\mu)_{1,3}^{3,1} ( e_{1,3}\otimes e_{3,1} + e_{3,1}\otimes e_{1,3})
\nonumber \\
&+&
R(\lambda,\mu)_{1,4}^{4,1} ( e_{1,4}\otimes e_{4,1} + e_{4,1}\otimes e_{1,4})
+
R(\lambda,\mu)_{1,4}^{3,2} ( e_{1,3}\otimes e_{4,2} + e_{3,1}\otimes e_{2,4})
\nonumber \\
&+&
R(\lambda,\mu)_{1,4}^{2,3} ( e_{1,2}\otimes e_{4,3} + e_{2,1}\otimes e_{3,4})
+
R(\lambda,\mu)_{1,4}^{1,4} e_{1,1}\otimes e_{4,4}
+
R(\lambda,\mu)_{2,1}^{2,1} e_{2,2}\otimes e_{1,1}
\nonumber \\
&+&
R(\lambda,\mu)_{2,2}^{3,1} ( e_{2,3}\otimes e_{2,1} + e_{3,2}\otimes e_{1,2})
+
R(\lambda,\mu)_{2,2}^{2,2} e_{2,2}\otimes e_{2,2}
\nonumber \\
&+&
R(\lambda,\mu)_{2,3}^{2,3} e_{2,2}\otimes e_{3,3} 
+
R(\lambda,\mu)_{2,3}^{3,2} e_{2,3} \otimes e_{3,2} 
+
R(\lambda,\mu)_{3,2}^{2,3} e_{3,2}\otimes e_{2,3} 
+
R(\lambda,\mu)_{3,2}^{3,2} e_{3,3} \otimes e_{2,2} 
\nonumber \\
&+&
R(\lambda,\mu)_{3,1}^{3,1} e_{3,3} \otimes e_{1,1}
+
R(\lambda,\mu)_{3,3}^{3,3} e_{3,3} \otimes e_{3,3}
+
R(\lambda,\mu)_{4,1}^{4,1} e_{4,4} \otimes e_{1,1}
+
R(\lambda,\mu)_{4,2}^{4,2} e_{4,4} \otimes e_{2,2}
\nonumber \\
&+&
R(\lambda,\mu)_{2,4}^{2,4} e_{2,2} \otimes e_{4,4}
+
R(\lambda,\mu)_{2,3}^{4,1} ( e_{2,4}\otimes e_{3,1} + e_{4,2}\otimes e_{1,3})
\nonumber \\
&+&
R(\lambda,\mu)_{2,4}^{3,3} ( e_{2,3}\otimes e_{4,3} + e_{3,2}\otimes e_{3,4})
+
R(\lambda,\mu)_{2,4}^{4,2} ( e_{2,4}\otimes e_{4,2} + e_{4,2}\otimes e_{2,4})
\nonumber \\
&+&
R(\lambda,\mu)_{3,2}^{4,1} ( e_{3,4}\otimes e_{2,1} + e_{4,3}\otimes e_{1,2})
+
R(\lambda,\mu)_{3,3}^{4,2} ( e_{3,4}\otimes e_{3,2} + e_{4,3}\otimes e_{2,3})
\nonumber \\
&+&
R(\lambda,\mu)_{3,4}^{4,3} ( e_{3,4}\otimes e_{4,3} + e_{4,3}\otimes e_{3,4})
+
R(\lambda,\mu)_{3,4}^{3,4} ( e_{3,3}\otimes e_{4,4} - e_{4,4}\otimes e_{3,3}).
\ear

The corresponding amplitudes $R(\lambda,\mu)_{a,b}^{c,d}$ are given by,
\bear
R(\lambda,\mu)_{1,1}^{1,1}&=&(1-\mu \lambda)(1-\mu \lambda w)(1-\mu \lambda w^2),
\\
R(\lambda,\mu)_{1,2}^{1,2}&=&(\lambda-\mu)(1-\mu \lambda w)(1-\mu \lambda w^2),
\\
R(\lambda,\mu)_{1,2}^{2,1}&=& \sqrt{(1-\mu^2)(1-\lambda^2)} (1-\mu \lambda w) (1-\mu \lambda w^2),
\\
R(\lambda,\mu)_{1,3}^{1,3}&=&(\lambda-\mu)(\lambda-\mu w) (1-\mu \lambda w^2),
\\
R(\lambda,\mu)_{1,3}^{2,2}&=&\sqrt{(1-\lambda^2)(1-\mu^2 w)(1+w)} (\lambda-\mu) (1-\mu \lambda w^2),
\\
R(\lambda,\mu)_{1,3}^{3,1}&=& \sqrt{(1-\lambda^2)(1-\lambda^2 w)(1-\mu^2)(1-\mu^2 w)} (1-\mu \lambda w^2),
\ear
\bear
R(\lambda,\mu)_{2,2}^{2,2}&=&((1-\lambda^2)(1-\mu^2 w)-(\mu-\lambda)(\mu-\lambda w)) (1-\mu \lambda w^2),
\\
R(\lambda,\mu)_{2,2}^{3,1}&=&(\mu-\lambda) \sqrt{(1-\mu^2)(1-\lambda^2 w)(1+w)} (1-\mu \lambda w^2),
\\
R(\lambda,\mu)_{2,3}^{2,3}&=&(\lambda-\mu) ((1-\mu^2 \lambda^2)(1-w^3)-w (\lambda-\mu w) (\lambda-\mu)),
\\
R(\lambda,\mu)_{2,3}^{3,2}&=& \sqrt{(1-\mu^2 w)(1-\lambda^2 w)} ( (1-\mu^2) (1-\lambda^2 w^2)-(1+w)(\lambda-\mu w) (\lambda- \mu)),
\\
R(\lambda,\mu)_{3,1}^{3,1}&=&(\mu-\lambda)(\mu-\lambda w)(1-\lambda \mu w^2),
\\
R(\lambda,\mu)_{1,4}^{4,1}&=& \sqrt{(1-\lambda^2) (1-\lambda^2 w) (1-\lambda^2 w^2)} \sqrt{(1-\mu^2) (1-\mu^2 w) (1-\mu^2 w^2)},
\\
R(\lambda,\mu)_{1,4}^{3,2}&=& \sqrt{(1-\lambda^2) (1-\lambda^2 w)} \sqrt{(1-\mu^2 w) (1-\mu^2 w^2)} \sqrt{1+w+w^2} (\lambda-\mu),
\\
R(\lambda,\mu)_{1,4}^{2,3}&=& \sqrt{(1-\lambda^2) (1-\mu^2 w^2)} \sqrt{1+w+w^2} (\lambda-\mu) (\lambda- \mu w),
\\
R(\lambda,\mu)_{1,4}^{1,4}&=&(\lambda-\mu)(\lambda-\mu w)(\lambda-\mu w^2),
\\
%R(\lambda,\mu)_{2,1}^{2,1}&=&(\mu-\lambda)(1-\lambda \mu w)(1-\lambda \mu w^2)
%\\
R(\lambda,\mu)_{2,3}^{4,1}&=& \sqrt{(1-\lambda^2 w) (1-\lambda^2 w^2)} \sqrt{(1-\mu^2) (1-\mu^2 w)} \sqrt{1+w+w^2} (\mu-\lambda),
\\
R(\lambda,\mu)_{2,4}^{4,2}&=& \sqrt{(1-\lambda^2 w) (1-\lambda^2 w^2)} \sqrt{(1-\mu^2 w) (1-\mu^2 w^2)} (1-\mu \lambda),
\\
R(\lambda,\mu)_{2,4}^{3,3}&=& \sqrt{(1-\lambda^2 w) (1-\mu^2 w^2)} \sqrt{(1+w)(1+w+w^2)} (1-\lambda \mu)(\lambda-\mu),
\\
R(\lambda,\mu)_{2,4}^{2,4}&=& (1+w+w^2) (1-\lambda \mu)(\lambda-\mu)(\lambda-\mu w),
\\
R(\lambda,\mu)_{3,2}^{3,2}&=&((1-\lambda^2 \mu^2)(1+w)- w(\mu-\lambda)(\mu-\lambda w)) (\mu - \lambda),
\\
R(\lambda,\mu)_{3,2}^{2,3}&=&((1-\lambda^2)(1-\mu^2 w^2)-(1+w)(\mu-\lambda)(\mu-\lambda w)) \sqrt{(1-\lambda^2 w) (1-\mu^2 w)}, 
\\
R(\lambda,\mu)_{3,2}^{4,1}&=& \sqrt{(1-w^2 \lambda^2)(1-\mu^2)} \sqrt{1+w+w^2} (\mu-\lambda)(\mu-\lambda w),
\\
R(\lambda,\mu)_{3,3}^{4,2}&=& \sqrt{(1-\lambda^2 w^2) (1-\mu^2 w)} \sqrt{(1+w)(1+w+w^2)} (1-\lambda \mu)(\mu-\lambda),
\\
R(\lambda,\mu)_{3,3}^{3,3}&=& ((1-\lambda^2 w)(1-\mu^2 w^2)-(1+w+w^2)(\mu-\lambda)(\mu-\lambda w)) (1-\lambda \mu),
\\
R(\lambda,\mu)_{3,4}^{4,3}&=& \sqrt{(1-\lambda^2 w^2) (1-\mu^2 w^2)} (1-\lambda \mu) (1-\lambda \mu w),
\\
R(\lambda,\mu)_{3,4}^{3,4}&=& (1+w+w^2) (1-\lambda \mu) (1-\lambda \mu w)(\lambda-\mu),
\\
R(\lambda,\mu)_{4,1}^{4,1}&=& (\mu-\lambda) (\mu-\lambda w) (\mu-\lambda w^2),
\\
R(\lambda,\mu)_{4,2}^{4,2}&=& (1+w+w^2)(1-\lambda \mu) (\mu-\lambda) (\mu-\lambda w),
\ear
where $\omega =\pm \imath$.

{\section*{\bf Appendix C : $SL(2,{\cal R})$ $R$-matrix elements}
\setcounter{equation}{0}
\renewcommand{\theequation}{C.\arabic{equation}}

Here we present the amplitudes $R(\lambda,\mu)_{a,b}^{c,d}$ corresponding to 
the submatrices of the $R$-matrix for $SL(2,{\cal R})$.
The results are given up to the particle sector $n=4$,
\begin{itemize}
\item $n=0$
\EQ
R(\lambda,\mu)_{1,1}^{1,1}=1.
\EN

\item $n=1$
\bear
R(\lambda,\mu)_{1,2}^{1,2}
&=&
R(\lambda,\mu)_{2,1}^{2,1}
=
\frac{\imath (\mu-\lambda)}{p_2(\lambda,\mu)},
\\
R(\lambda,\mu)_{1,2}^{2,1}
&=&
R(\lambda,\mu)_{2,1}^{1,2}
=
\frac{2 {\bf s}}{p_2(\lambda,\mu)}.
\ear

\item $n=2$
\bear
R(\lambda,\mu)_{1,3}^{1,3}
&=&
R(\lambda,\mu)_{3,1}^{3,1}
=
-\frac{(\mu-\lambda) (\imath+\mu-\lambda)}{p_3(\lambda,\mu)},
\\
R(\lambda,\mu)_{1,3}^{2,2}
&=&
R(\lambda,\mu)_{2,2}^{1,3}
=
R(\lambda,\mu)_{2,2}^{3,1}
=
R(\lambda,\mu)_{3,1}^{2,2}
=
\frac{2 \sqrt{\imath{\bf s}} \sqrt{\imath (2 {\bf s}-1)} (\mu-\lambda)}{p_3(\lambda,\mu)},
\\
R(\lambda,\mu)_{1,3}^{3,1}
&=&
R(\lambda,\mu)_{3,1}^{1,3}
=
\frac{2 {\bf s}(2 {\bf s}-1)}{p_3(\lambda,\mu)},
\\
R(\lambda,\mu)_{2,2}^{2,2}
&=&
\frac{2 {\bf s}(2 {\bf s}-1)-(\mu-\lambda)(-\imath+\mu-\lambda)}{p_3(\lambda,\mu)}.
\ear

\item $n=3$
\bear
R(\lambda,\mu)_{1,4}^{1,4}
&=&
R(\lambda,\mu)_{4,1}^{4,1}
=
\frac{(2-\imath\mu+\imath \lambda) (\mu-\lambda) (\imath+\mu-\lambda)}{p_4(\lambda,\mu)},
\\
R(\lambda,\mu)_{1,4}^{2,3}
&=&
R(\lambda,\mu)_{2,3}^{1,4}
=
R(\lambda,\mu)_{3,2}^{4,1}
=
R(\lambda,\mu)_{4,1}^{3,2}
\nonumber \\
&=&
\frac{2 \imath \sqrt{3} \sqrt{\imath(-1+{\bf s})} \sqrt{\imath {\bf s}} (\mu-\lambda) (\imath+\mu-\lambda)}{p_4(\lambda,\mu)},
\\
R(\lambda,\mu)_{1,4}^{3,2}
&=&
R(\lambda,\mu)_{2,3}^{4,1}
=
R(\lambda,\mu)_{3,2}^{1,4}
=
R(\lambda,\mu)_{4,1}^{2,3}
\nonumber \\
&=&
\frac{2 \sqrt{3} \sqrt{\imath(1-{\bf s})} \sqrt{\imath {\bf s}} (-1+2 {\bf s}) (\mu-\lambda) }{p_4(\lambda,\mu)},
\ear
\bear
%\\
R(\lambda,\mu)_{1,4}^{4,1}
&=&
R(\lambda,\mu)_{4,1}^{1,4}
=
\frac{4 (-1+ {\bf s}) {\bf s} (-1+ 2 {\bf s})}{p_4(\lambda,\mu)},
\\
R(\lambda,\mu)_{2,3}^{2,3}
&=&
R(\lambda,\mu)_{3,2}^{3,2}
=
\frac{-\imath (\mu-\lambda) \left[-8 (-1+{\bf s}) {\bf s}+(\mu-\lambda)(-\imath+\mu-\lambda) \right]} {p_4(\lambda,\mu)},
\\
R(\lambda,\mu)_{2,3}^{3,2}
&=&
R(\lambda,\mu)_{3,2}^{2,3}
=
\frac{2(-1+2 {\bf s}) \left[ 2(-1+{\bf s}){\bf s} - (\mu-\lambda)(-\imath+\mu-\lambda) \right]}{p_4(\lambda,\mu)}.
\ear

\item $n=4$
\bear
R(\lambda,\mu)_{1,5}^{1,5}
&=&
R(\lambda,\mu)_{5,1}^{5,1}
=
\frac{(\mu-\lambda) (\imath+\mu-\lambda) (2 \imath+\mu-\lambda) (3 \imath +\mu-\lambda)}{p_5(\lambda,\mu)},
\\
R(\lambda,\mu)_{1,5}^{2,4}
&=&
R(\lambda,\mu)_{2,4}^{1,5}
=
R(\lambda,\mu)_{4,2}^{5,1}
=
R(\lambda,\mu)_{5,1}^{4,2}
\nonumber \\
&=&
-\frac{2 \sqrt{2} \sqrt{\imath {\bf s}} \sqrt{\imath(-3+2 {\bf s})} (\mu-\lambda) (\imath+\mu-\lambda) (2 \imath+\mu-\lambda) }{p_5(\lambda,\mu)},
\\
R(\lambda,\mu)_{1,5}^{3,3}
&=&
R(\lambda,\mu)_{3,3}^{1,5}
=
R(\lambda,\mu)_{3,3}^{5,1}
=
R(\lambda,\mu)_{5,1}^{3,3}
\nonumber \\
&=&
\frac{2 \sqrt{6} \sqrt{\imath (-1+{\bf s})} \sqrt{\imath {\bf s}} \sqrt{\imath (-3+2 {\bf s})} \sqrt{\imath (-1+2 {\bf s})} (\mu-\lambda) (\imath+\mu-\lambda) }{p_5(\lambda,\mu)},
\\
R(\lambda,\mu)_{1,5}^{4,2}
&=&
R(\lambda,\mu)_{2,4}^{5,1}
=
R(\lambda,\mu)_{4,2}^{1,5}
=
R(\lambda,\mu)_{5,1}^{2,4}
\nonumber \\
&=&
\frac{4 \sqrt{2} (-1+{\bf s}) \sqrt{\imath {\bf s}} \sqrt{\imath (-3+2 {\bf s})} (-1+2 {\bf s}) (\mu-\lambda) }{p_5(\lambda,\mu)},
\\
R(\lambda,\mu)_{1,5}^{5,1}
&=&
R(\lambda,\mu)_{5,1}^{1,5}
=
\frac{4 (-1+{\bf s}) {\bf s} (-3+2{\bf s}) (-1+2 {\bf s}) }{p_5(\lambda,\mu)},
\\
R(\lambda,\mu)_{2,4}^{2,4}
&=&
R(\lambda,\mu)_{4,2}^{4,2}
=
\frac{(\mu-\lambda) (\imath+\mu-\lambda) \left[ 6(3-2 {\bf s}){\bf s} +(\mu-\lambda) (-\imath+\mu-\lambda) \right] }{p_5(\lambda,\mu)},
\\
R(\lambda,\mu)_{2,4}^{3,3}
&=&
R(\lambda,\mu)_{3,3}^{2,4}
=
R(\lambda,\mu)_{4,2}^{3,3}
=
R(\lambda,\mu)_{3,3}^{4,2}
\nonumber \\
&=&
\frac{2 \sqrt{3} \sqrt{\imath(-1+{\bf s})} \sqrt{\imath(-1+2 {\bf s})} (\mu-\lambda) \left[ 2 (-3+2 {\bf s}) {\bf s} -(\mu-\lambda) (-\imath+\mu-\lambda) \right] }{p_5(\lambda,\mu)},
\nonumber \\
\\
R(\lambda,\mu)_{2,4}^{4,2}
&=&
R(\lambda,\mu)_{4,2}^{2,4}
=
\frac{2 (-1+{\bf s}) (-1+2 {\bf s}) \left[ 2 (-3+2 {\bf s}) {\bf s} - 3 (\mu-\lambda) (-\imath+\mu-\lambda) \right] }{p_5(\lambda,\mu)},
\nonumber \\
\\
R(\lambda,\mu)_{3,3}^{3,3}
&=&
\frac{4 (-1+{\bf s}) {\bf s} (-3+2 {\bf s}) (-1+2 {\bf s}) +2 \imath \left[ 3+4{\bf s}(-3+2 {\bf s}) \right] (\mu-\lambda)
}{p_5(\lambda,\mu)}
\nonumber \\
&+& \frac{-\left[ 7+8 {\bf s} (-3+2 {\bf s}) \right] (\mu-\lambda)^2 -2 \imath (\mu-\lambda)^3 +(\mu-\lambda)^4}{p_5(\lambda,\mu)}.
\ear
\end{itemize}

The auxiliary function 
$p_i(\lambda,\mu)$ entering in the above expressions is defined by,
\EQ
p_i(\lambda,\mu)
=
\prod_{j=1}^{i-1}[
2 {\bf s}+1 + \imath (\mu-\lambda)-j].
\EN

\section*{\bf Appendix D : The coordinate Bethe ansatz}
\setcounter{equation}{0}
\renewcommand{\theequation}{D.\arabic{equation}}

In this appendix we present the coordinate Bethe ansatz diagonalization of the non-compact Hamiltonian,
\EQ
H({\bf s}) \psi_n = E_n \psi_n
\label{eig1}
\EN
up to the two-particle sector $n=2$.

Considering Eqs.(\ref{hnon},\ref{hnon1}) the action of the Hamiltonian on 
the subspace of states up to the sector $n=2$ are,
\bear 
H_{12}({\bf s}) \ket{0,0} &=&0 \\ 
H_{12}({\bf s}) \ket{0,1} &=& \ket{0,1}-\ket{1,0} \\
H_{12}({\bf s}) \ket{1,0} &=& \ket{1,0}-\ket{0,1} \\
H_{12}({\bf s}) \ket{1,1} &=& 2 \ket{1,1}+d({\bf s}) (\ket{0,2}+\ket{2,0}) \\
H_{12}({\bf s}) \ket{2,0} &=& e({\bf s}) \ket{0,2}+d({\bf s}) \ket{1,1}+c({\bf s})\ket{2,0} \\
H_{12}({\bf s}) \ket{0,2} &=& c({\bf s}) \ket{0,2}+d({\bf s}) \ket{1,1}+e({\bf s})\ket{2,0}
\ear
where the parameters $c({\bf s})$, $d({\bf s})$ and $e({\bf s})$ are given by,
\EQ
c({\bf s})=2+\frac{1}{2 {\bf s}-1}, ~~
d({\bf s})=-\frac{2 \sqrt{\bf s}}{\sqrt{2 {\bf s}-1}}, ~~
e({\bf s})=\frac{1}{2 {\bf s}-1}
\EN

The sector $n=0$ only contains the reference state and the wave-function is $\psi=\ket{0 \dots 0}$ with $E_0=0$.
The wave-function $\psi_1$ for the $n=1$ sector is the linear combination of 
states $\ket{0 \dots \substack{\underbrace{1} \\ x} \dots 0}$ made by inserting one excitation of type $1$ on a $x$-th site,
\EQ
\psi_1
=
\sum_{x=1}^L \phi(x) \ket{0 \dots \substack{\underbrace{1} \\ x} \dots 0}
\EN

The action of $H({\bf s})$ on $\psi_1$ leads us to the following difference equation for $\phi(x)$
\EQ
E_1 \phi(x)=2 \phi(x)- \phi(x+1)-\phi(x-1)
\label{dif1}
\EN

The standard plane-wave assumption $\phi(x)=\exp(\imath k x)$ solves Eq.(\ref{dif1}) provided the eigenvalue $E_1$ is
\EQ
E_1=2 \left[ 1 - \cos(k) \right]
\EN
while the one-particle momenta is fixed by periodic boundary condition $\phi(x+L)=\phi(x)$,
\EQ
\exp(\imath k L)=1
\EN

The subspace of states for $n=2$ is constituted of $\frac{L(L-1)}{2}$ states of two excitation of type 1
$\ket{0 \dots \substack{\underbrace{1} \\ x_1} \dots \substack{\underbrace{1} \\ x_2} \dots 0}$ 
and $L$ states of one excitation of type $2$  $\ket{0 \dots \substack{\underbrace{2} \\ x} \dots 0}$.
Therefore, the ansatz for the two-particle wave-function $\psi_2$ is,
\EQ
\psi_2
=
\sum_{1 \le x_1 < x_2 \le L} \phi(x_1,x_2)
\ket{0 \dots \substack{\underbrace{1} \\ x_1} \dots \substack{\underbrace{1} \\ x_2} \dots 0}
+
\sum_{x=1}^{L} \varphi(x)
\ket{0 \dots \substack{\underbrace{2} \\ x} \dots 0}.
\EN

By solving the eigenvalue equation (\ref{eig1}) for $\psi_2$ we derive the following set of difference equations,
\bear
&\mbox{\textbullet}& (E-4) \phi(x_1,x_2)=-\phi(x_1+1,x_2)-\phi(x_1-1,x_2)-\phi(x_1,x_2+1)-\phi(x_1,x_2-1), ~~x_2>x_1+1 \label{dif21} \nonumber \\
\\
&\mbox{\textbullet}& (E-4) \phi(x_1,x_2)=-\phi(x_1,x_2+1)-\phi(x_1-1,x_2)+d({\bf s}) \left[ \varphi(x_1)+\varphi(x_2) \right], ~~x_2=x_1+1 \label{dif22} \\
&\mbox{\textbullet}& (E-2 c) \varphi(x)=d({\bf s}) \left[ \phi(x-1,x)+\phi(x,x+1) \right] + e({\bf s}) \left[ \varphi(x-1)+\varphi(x+1) \right], \label{dif23} \ear

To solve Eqs.(\ref{dif21}-\ref{dif23}) we have to consider three basic steps.
First, guided by the $n=1$ case and the structure of (\ref{dif21}), we proposed for
$\phi(x_1,x_2)$ the typical Bethe ansatz form, 
\EQ
\phi(x_1,x_2)= \exp(\imath k_1 x_1 + \imath k_2 x_2) + S(k_1,k_2) \exp(\imath k_2 x_1 + \imath k_1 x_2)
\label{bea21}
\EN
where $S(k_1,k_2)$ is the two-particle scattering matrix.
This solves Eq.(\ref{dif21}) provided the eigenvalue is
\EQ
E_2 = 2 \left[1 - \cos(k_1) \right] + 2 \left[1 - \cos(k_2) \right].
\EN

Next we consider the compatibility between Eq.(\ref{dif22}) and Eq.(\ref{dif23}) at the point $x_2=x_1+1$.
By subtracting Eq.(\ref{dif22}) from Eq.(\ref{dif21}) 
for $x_1=x$ and $x_2=x+1$ we obtain the following matching condition,
\EQ
d({\bf s}) \varphi(x) = - \phi(x,x).
\label{bea22}
\EN

Now by substituting the ansatz Eq.(\ref{bea21}) in Eq.(\ref{dif23}) after 
considering condition (\ref{bea22}) we are able to fix the scattering matrix as,
\EQ
S(k_1,k_2)
=
-
\frac{1+\exp\left[\imath (k_1+k_2) \right] + (2{\bf s} - 1) \exp\left[\imath k_1 \right] - (2 {\bf s}+1) \exp\left[\imath k_2 \right]}
{1+\exp\left[\imath (k_1+k_2) \right] + (2{\bf s} - 1) \exp\left[\imath k_2 \right] - (2 {\bf s}+1) \exp\left[\imath k_1 \right]}.
\EN

The final step is to impose the periodic boundary conditions $\phi(x_1,x_2+L)=\phi(x_2,x_1)$.
As a result we obtain the two-particle Bethe ansatz constraint,
\EQ
\exp(\imath k_i L) \prod_{\stackrel{j=1}{j \ne i}}^{2} S(k_i,k_j)=1,
\EN
where we have used the unitarity property $S(k_1,k_2) S(k_2,k_1)=1$.

In order to reproduce the Bethe ansatz equations (\ref{beaXXX}) we have
to parameterize the momenta $k_j$ as,
\EQ
\exp(\imath k_j)= \frac{\lambda_j +2{\bf s} \imath}{\lambda_j}. 
\EN

\end{document}